\documentclass{bmcart}

\usepackage{amsthm,amsmath}
\usepackage[applemac]{inputenc} 

\usepackage{amsfonts,amsmath}

\usepackage{graphicx}

\startlocaldefs

\def\ds{\displaystyle}

\def\({ \biggr (   }
\def\){ \biggr ) }

\newcommand{\e}{{\rm e}}

\newcommand{\R}{{\mathbf R}}
\newcommand{\C}{{\mathbf C}}

\newcommand{\rr}{{\mathbf r}}
\newcommand{\ww}{{\mathbf w}}

\newcommand{\pp}{{\mathbf p}}

\renewcommand{\theequation}{\thesection.\arabic{equation}}

\usepackage{color}
\definecolor{darkred}{rgb}{.91, .26, .26}

\endlocaldefs

\begin{document}

\begin{frontmatter}

\begin{fmbox}     
\dochead{Research}



\title{Laminar model for the joint development of ocular dominance columns and CO blobs in V1}


\author[
   addressref={aff1},                   
   corref={aff1},                       
   email={aoster@ewu.edu}   
]{\inits{AM}\fnm{Andrew M.} \snm{Oster}}


\address[id=aff1]{
  \orgname{Department of Mathematics, Eastern Washington University}, 
  \street{Kingston Hall},
    \city{Cheney,  WA},                              
     \postcode{99004},    
     \cny{USA}                                    
}


\begin{artnotes}
\end{artnotes}

\end{fmbox}


\begin{abstractbox}
\begin{abstract} 

In this paper, we present a multi--layer, activity--dependent model for the joint development of ocular dominance (OD) columns and cytochrome oxidase (CO) blobs in primate V1. For simplicity, we focus on layers 4C and 2/3 with both layers receiving direct thalamic inputs and layer 4C sending vertical projections to layer 2/3. Both the thalamic and the vertical connections are taken to be modifiable by activity. Using a correlation--based Hebbian learning rule with subtractive normalization, we show how the formation of an OD map in layer 4C is inherited by layer 2/3 via the vertical projections. Competition between these feedforward projections and the direct thalamic input to layer 2/3 then results in the formation of CO blobs superimposed upon the ocular dominance map. The spacing of the OD columns is determined by the spatial profile of the intralaminar connections within layer 4, while the spacing of CO blobs depends both on the width of the OD columns inherited from layer 4 and the spatial distribution of intralaminar connections within the superficial layer. The resulting CO blob distribution is shown to be consistent with experimental data.  In addition, we numerically simulate monocular deprivation and find that while the CO blob distribution is unaltered, the OD pattern undergoes modification.  The OD stripes of the deprived eye narrow, whereas the OD stripes for the remaining open eye widen.


\end{abstract}




\end{abstractbox}
%

\end{frontmatter}



\section{Introduction}

Most neurons in the primary visual cortex (V1) respond preferentially to a variety of features associated with visual stimuli, including orientation, ocular dominance (OD), spatial frequency, and direction selectivity.  As one progresses tangentially across V1, the response properties vary in a nearly continuous fashion.   Seminal studies by Hubel \& Wiesel \cite{Hubel:1962:FAC, Hubel:1974:SRG} revealed the existence of ocularity and orientation preference feature maps.  Furthermore, they found that neurons through the cortical layers have similar response properties suggesting that the visual cortex is organized in a columnar fashion. It was further conjectured that the feature preferences of cortical neurons are generated by the convergence of thalamic afferents on to input layer 4, and then passed on to other layers through vertical interlaminar projections. It follows from this that the formation of feature preference maps can be understood in terms of the development of feedforward connections from thalamus to layer 4. The latter has been the focus of most models of cortical development \cite{Swindale:1996:DTV, vanOoyen:2001:REV}. 

Although the columnar hypothesis proposed by Hubel and Wiesel may be approximately true in cat, it is clearly an oversimplification in primates such as the macaque, where feature preferences are not uniform throughout the depth of a column \cite{ LeVay:1991:COV, Lund:2003:ASF}. For example, there are two major classes of thalamic cells projecting to input layer 4C of the macaque monkey, magnocellular (M) and parvocellular (P). The M and P pathways connect to layer 4C in a graded fashion with the M pathway dominating upper 4C$\alpha$ and the P pathway dominating lower 4C$\beta$ \cite{Levitt:1996:ASE, Callaway:1998:LCV}.  Interestingly, orientation--selective cells within layer 4C also appear to emerge in a graded fashion, being found predominantly in mid and upper layer 4C$\alpha$ \cite{Blasdel:1984:POM, Hawken:1984:OSL}, and coinciding with the emergence of horizontal axon projections from excitatory spiny stellate cells \cite{Lund:1987:LCN,Yoshioka:1994:TCM}. It is possible that such projections, rather than direct thalamic inputs, provide the anatomical substrate for the generation of orientation preference \cite{Adorjan:1999:MIO}. A second example arises from the observation that direction--selective cells first emerge in layer 4B but are not found in superficial layers 2/3 \cite{Hawken:1988:LOD}. 

Yet another important example of nonuniformity through the layers is the occurrence of cytochrome oxidase (CO) blobs in superficial layers of macaque V1 \cite{Horton:1981:COB, Horton:1984:COP}. These regions of higher metabolic activity receive direct inputs from a third class of thalamic cells, namely those forming the koniocellular (K) pathway \cite{ Casagrande:1994:KKK,Hendry:2000:KPV}. 
The spatial distribution of CO blobs within cortex is correlated with a number of
feature preference maps. For example,
the blobs are found at evenly spaced intervals along the center of OD
columns \cite{Horton:1984:COP}, and neurons within the blobs tend to be less binocular and less orientation
selective \cite{Livingstone:1984:SIC}. The blobs are also linked with low spatial frequency domains
\cite{Tootell:1988:FAM}    and appear to be coincident with singularities in the orientation preference map  \cite{Obermayer:1993:GOO}. 
The arrangement of CO blobs is
 reflected anatomically by the distribution of intrinsic horizontal connections, which tend to
link blobs-to--blobs and interblobs--to--interblobs \cite{Livingstone:1984:SIC, Yoshioka:1996:RPL, Yabuta:1998:COB}, and by extrinsic corticocortical connections linking blobs to specific compartments in V2 and other extrastriate areas
\cite{Livingstone:1984:SIC, Sincich:2002:DCO}. Taken together these observations suggest that the CO blobs are sites
of functionally and anatomically distinct channels of visual processing. Note that CO blobs have also been found in cat, but tend to exhibit weaker functional properties  \cite{Murphy:1995:COB, Shoham:1997:STF}. For example, although blobs avoid OD borders in cat visual cortex, they are not strictly aligned along the center of the columns. 

From a developmental perspective, the fact that cells in different layers of primate V1 may have different sets of feature preferences and receive distinct classes of thalamic input, suggests that one should consider a multi-layer cortical model. Such a model would first need to explain how direct thalamic inputs to different layers develop certain common features. For example, how do the OD maps of the M,P and K pathways align with each other? Second, it would need to account for the emergence of layer specific features as exemplified by the CO blobs. In this paper we construct such a laminar network model in order to investigate the joint development of OD columns and CO blobs in primate V1. For simplicity, we focus on layers 4C and 2/3. Both layers receive direct thalamic inputs and layer 4C sends vertical projections to layer 2/3. We assume that the OD map first develops in layer 4C and that this is then inherited by the thalamic inputs to layer 2/3 via the vertical projections from layer 4C.  We also assume that both the thalamic and the vertical connections are modifiable by activity during the early development of OD columns. Using a correlation--based Hebbian learning rule with subtractive normalization \cite{Miller:1989:ODC}, we show how the feedforward vertical projections from layer 4C to layer 2/3 compete with the direct thalamic input to layer 2/3, resulting in the formation of CO blobs superimposed upon the OD map. As in previous activity--based models \cite{Willshaw:1976:PNC,Swindale:1980:MFO,Miller:1989:ODC}, the spacing of the OD columns is determined by the spatial profile of the intralaminar connections within layer 4. This is consistent with recent experimental findings by Hensch and Stryker \cite{Hensch:2004:INH}, who show that altering the lateral inhibitory circuits in kitten visual cortex during development results in a widening or narrowing of the ocular dominance columns. The spacing of CO blobs, on the other hand, depends both on the width of the OD columns inherited from layer 4 and the spatial distribution of intralaminar connections within superficial layer 2/3.

We analyze our multi--layer developmental model by carrying out a perturbation expansion with respect to the strength of coupling of the intralaminar connections. Such a perturbation expansion is implicit in standard single-layer correlation-based Hebbian models  \cite{Miller:1989:ODC, Swindale:1996:DTV}), in which the steady-state activity within the layer is obtained by inverting a nonlocal operator that depends on the distribution of intralaminar connections. The resulting inverse operator consists of a convolution with a modified interaction kernel, which is usually taken to be a Mexican hat function. Such an interaction kernel generates a spatially correlated competition between left and right eye inputs that ultimately leads to the formation of OD columns. One of the problems with taking a Mexican hat function, however, is that the convolution kernel is not invertible. On the other hand preserving invertibility means that the OD pattern becomes more sensitive to initial conditions. As we show in this paper, such issues become even more significant within the context of a multi-layer network model. Another feature that emerges from our analysis is that it is necessary to use an adaptive subtractive normalization scheme in order to preserve locally the total synaptic density throughout the developmental process. That is, the form of the subtractive normalization changes whenever one of the weight components vanishes. 

It should be noted that a developmental model of the laminar structure of V1 has also been constructed by Grossberg and Williamson \cite{Grossberg:2001:ICD}. However, their model is concerned with the development  of the feedforward, horizontal and feedback circuits within V1 that are thought to be involved in perceptual grouping, attention and learning in adult cortex. They do not address the issue of how cortical feature maps develop across different layers. The joint development of OD columns and CO blobs has previously been studied by Nakagama and
Tanaka \cite{Tanaka:2004:COB}. They consider an abstract thermodynamic model for the joint development of CO blobs and OD columns, in which layers 2/3 and 4C are collapsed into a single
effective layer. Although their model successfully generates patterns that are consistent with experimental data, it is difficult to interpret the model directly in terms of the
various synaptic pathways from LGN to V1 and within V1. Moreover, the single layer model assumes that the development of the CO blobs mutually interacts with the development
of the OD columns, whereas in our more biologically realistic two--layer model the development of the OD columns influences the initial formation of the CO blobs but not
vice-versa.

\section{Developmental model}

\subsection{Laminar architecture of V1}

We begin by briefly describing the layered structure of macaque V1. Further details and references can be found in the reviews by Levitt, Lund and Yoshioka \cite{Levitt:1996:ASE} and Callaway
\cite{Callaway:1998:LCV}.  There are three major classes of LGN neurons projecting to the cortex: magnocellular (M), parvocellular (P), and koniocellular (K). Each class of LGN neuron receives input
from a different type of retinal ganglion cell and sends axons to distinct regions of V1. 

\noindent (i) The M pathway originates from retinal ganglion cells that innervate the two most
ventral M layers of the LGN. They have large receptive fields, are color insensitive, respond transiently to visual stimuli, prefer low spatial frequencies and are relatively
sensitive to luminance contrasts. Thus neurons in the M pathway are excellent for detecting subtle changes in luminance or rapidly moving stimuli, but are poorly suited for the
analysis of fine shape or color. The corresponding M-type LGN neurons project predominantly to layer 4C$\alpha$ with a weaker projection to layer 6. 

\noindent (ii) The P pathway originates from
retinal ganglion cells that innervate the four most dorsal $P$ layers of the LGN. They tend to be more numerous in the central retina and have smaller receptive fields than those
in the M pathway, allowing them to convey information on a finer spatial scale. They are also able to detect color contrast. On the other hand, P cells have a slower, more
sustained response to visual stimuli, making them less useful for the detection of rapid movement. The P-type LGN neurons project predominantly to layer 4C$\beta$ with weaker
projections to layers 4A and 6. 

\noindent (iii) The K pathway originates from the small retinal ganglion cells that innervate the thin intercalcated regions between the M and P
layers of macaque LGN. Until recently, these cells were thought to play a minor role in visual perception. However, there is a growing recognition of the importance of the
K pathway, in large part due to extensive studies of new world galago monkeys, whose
LGN has more distinct K layers than old world primates such as the macaque, thus making them more amenable to experimental analysis \cite{Casagrande:1994:KKK}. Across various primate
species one finds that K-type LGN cells have distinct neurochemical and receptive field properties, and specifically target the cytochrome oxidase (CO) blobs in layer 2/3 of cortex
\cite{Hendry:2000:KPV}. There are also some K cells that target layer 1. Although the functional role of the K pathway is still unclear, one likely contribution in diurnal
primates is color processing at short wavelengths.

\begin{figure}[htbp]
\begin{center}
\includegraphics[width=0.9\linewidth]{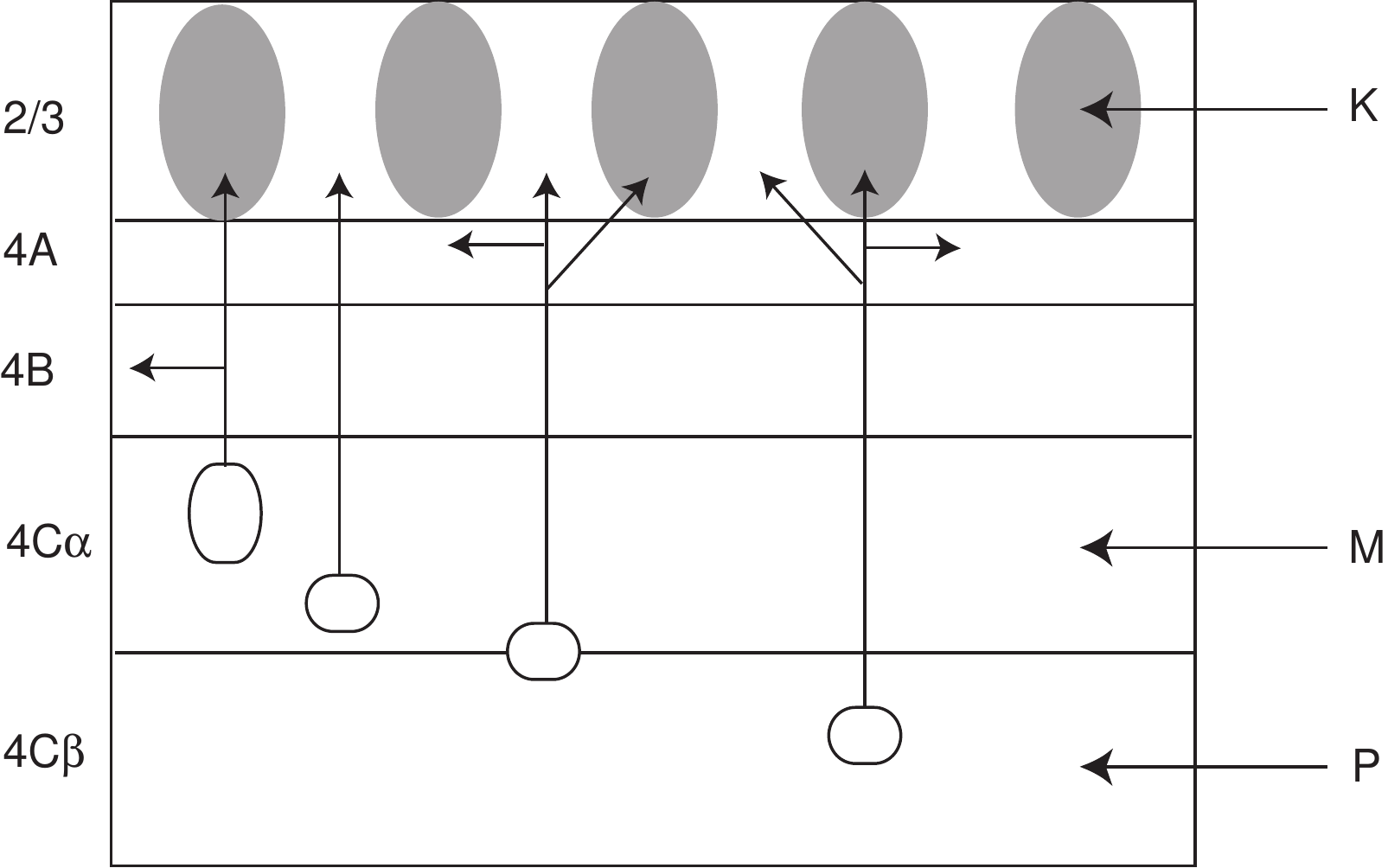}
\end{center}
\caption{\small Schematic diagram illustrating the main input pathways from LGN to V1 of macaque, and the major types of projections from layer 4C to superficial layer 2/3. See
text for details. Adapted from \cite{Callaway:1998:LCV}.}
\label{fig:MP}
\end{figure}

Spiny stellate neurons in layer 4C of macaque V1, which are the recipients of thalamocortical inputs, can be subdivided into at least four groups according to the location of
their dendritic fields within the layer and the targets of their projecting neurons in superficial layers \cite{Yoshioka:1996:RPL, Yabuta:1998:COB}. This is
illustrated in figure \ref{fig:MP}. The first group consists of neurons located in upper 4C$\alpha$ whose dendrites are confined to 4C$\alpha$ and whose axons arborize in
layers 4B and in the CO blobs of layer 2/3. The second group has narrowly stratified dendrites that are confined to lower 4C$\alpha$ and axons that target the interblobs in layer
2/3. Both of these neuron types are expected to belong to the M pathway. The third group straddles the border between 4C$\alpha$ and 4C$\beta$ and is thus likely to receive both M
and P afferents from the LGN, whereas the fourth group is found in layer 4C$\beta$ and has dendrites confined to layer 4C$\beta$. The axons of cells belonging to the last two
groups arborize in layers 4A and layer 2/3 blobs and interblobs. Note that there are some differences across primate species. For example, the interblob regions of diurnal
primates such as the squirrel monkey receive inputs only from P-type neurons in layer 4C$\beta$, whereas the interblob regions of nocturnal primates such as the galagos receive
inputs only from M-type neurons in layer 4C$\alpha$ \cite{Casagrande:1993:MAE}. In cats, W and Y cells of the LGN project directly to CO blobs whereas X cells project indirectly
via layer 4 to both blobs and interblobs \cite{Boyd:1996:LCG}.

\begin{figure}[htbp]
\begin{center}
\includegraphics[width=0.4\linewidth]{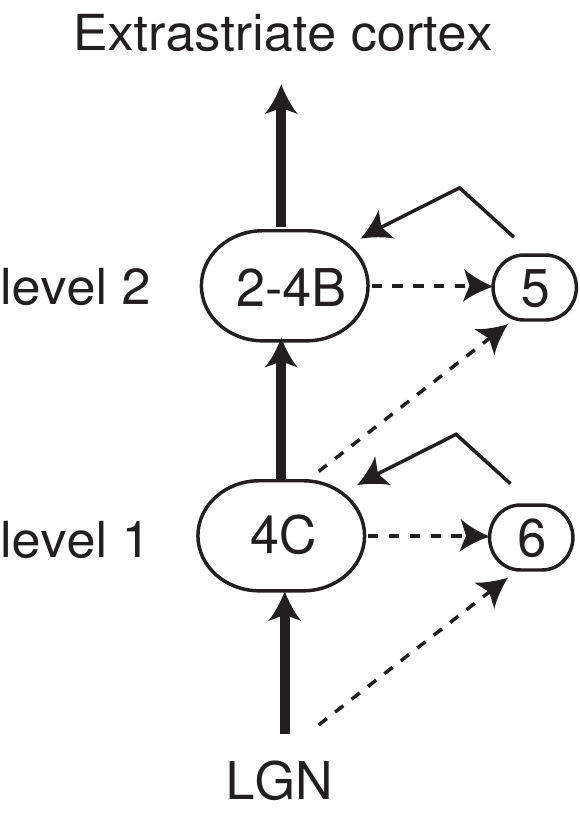}
\end{center}
\caption{\small Two--level
model of local V1 circuitry.  Adapted from \cite{Callaway:1998:LCV}.}
\label{fig:levels}
\end{figure}

Another important aspect of the laminar architecture of primary visual cortex is that each cortical layer sends its primary interlaminar output to
one other layer. This leads to a simplified two--level model of information processing by local circuits of V1 as illustrated in figure \ref{fig:levels} \cite{Callaway:1998:LCV}. Each level consists of one feedforward module and one feedback module. The feedforward modules (layer 4C at level 1 and layers 2-4B at level 2) receive strong excitatory
inputs from the previous level and make strong connections to the next higher level. The corresponding feedback modules (layer 6 at level 1 and layer 5 at level 2) receive
weaker input from the feedforward modules at the previous level and at the same level, and provide heavy feedback to the latter. It follows that the most direct path for
information flow from the LGN to extrastriate cortex is from LGN to layer 4C to layers 2-4B to layer 4 of extrastriate cortex. Note that the feedforward connections tend
to be focused whereas the feedback connections tend to be more diffuse in nature and play a modulatory role. A similar type of circuit exists in cat V1 if we consider layer 4C to
be analogous to cat's layer 4 and layers 2-4B analogous to layer 2/3.

\subsection{Reduced two--layer cortical model}

The laminar architecture outlined above is already present in newly born primates, as are ocular dominance columns and cytochrome oxidase blobs. The prenatal development of
OD columns is usually modeled in terms of a Hebbian-like competitive mechanism for the modification of left/right eye afferents from the LGN to layer 4C \cite{Willshaw:1976:PNC, Swindale:1980:MFO, Miller:1989:ODC, Swindale:1996:DTV, Harris:1997:CNT}.. In the case of primates, it is assumed that prenatal neural activity is
generated spontaneously either by retinal waves or by activity arising from the cortico-geniculate feedback loop \cite{Katz:2002:DCC}. Most developmental models do not
distinguish between sublayers 4C$\alpha$ and 4C$\beta$ nor the corresponding M and P pathways highlighted in figure \ref{fig:MP}. However, the observation that a subclass
of M-type neurons in upper layer 4C$\alpha$ innervate the interblobs, while LGN neurons from the K layers and a second subclass of M-type neurons in lower 4C$\alpha$ innervate the
blobs, suggests that there is some Hebbian--like competition between these two classes of afferents within layer 2/3. This motivates us to consider a two--layer cortical model for
the joint development of OD columns and CO blobs, which is consistent with the two--level architecture shown in figure \ref{fig:levels}. The basic assumptions of our model are as
follows:

\begin{enumerate}

\item[1.)] Ocular dominance columns in layer 4C are generated by an activity--dependent synaptic modification of feedforward afferents from the M and P layers of the LGN, involving
Hebbian--like competition between left and right eye afferents. The resulting pattern of OD columns is passed on to superficial layer 2/3 by vertical feedforward connections from
layer 4C to layer 2/3. 

\item[2.)] Cytochrome oxidase blobs in layer 2/3 are generated by an activity--dependent synaptic modification of feedforward afferents from the K layers of the LGN and
M-type feedforward afferents from layer 4C. This leads to an alternating pattern of blobs that are innervated by K-type inputs and one class of M-type inputs, and interblobs that
are innervated by a second class of M-type inputs. Since the drive from layer 4C is approximately monocular, the resulting pattern of CO blobs aligns with the OD columns through
the underlying Hebbian mechanism. The absence of feedback from layer 2/3 to layer 4C implies that the development of the CO blobs does not affect the formation of the OD
columns.

\end{enumerate}

\noindent Our model assumes that the vertical connections between layer 4C and layer 2/3 are modifiable by neural activity during the period when OD columns and CO blobs form.
There is some evidence for this in studies of cat visual cortex, where projections from layer 4C to layer 2/3 continue to mature until around postnatal week 6 whilst OD columns and
CO blobs begin to appear around postnatal week 3 \cite{Katz:1992:DLC}. Interestingly, vertical connections between other layers tend to mature much earlier. Indeed,
various {\em in vivo} and {\em in vitro} studies have shown that layer--specific intracortical connections in mammalian visual cortex tend to develop with a high degree of
specificity from the outset and are not the result of subsequent remodeling (see the reviews by Katz and Callaway \cite{Callaway:1990:ERC} and Bolz {\em et al.\ }\cite{Bolz:1996:SLC}). This suggests that growing
axons are able to distinguish specific cortical layers during development using activity--independent cues such as molecular markers. However, this picture is likely to be
oversimplified, since a number of recent studies indicate that activity--dependent mechanisms may regulate layer--specific axonal targeting for certain types of pyramidal neuron
targeting certain layers \cite{Hermann:1995:BAP, Callaway:1998:PDC, Butler01}.
Plasticity in vertical connections has also been found in experimental studies of development in the rodent somatosensory cortex  \cite{Feldman:2005:PSC}. Early in development (first postnatal week) whisker manipulations induce rapid changes in the whisker barrel map in layer 4C, consistent with plasticity of thalamic afferents. On the other hand, in older animals plasticity tends to occur first in layer 2/3 (and layer 5) and only later (if at all) in layer 4. A major cellular component of plasticity in layer 2/3 is deprivation--induced weakening of the layer 4 to layer 2/3 vertical projections via long--term depression \cite{Feldman:2003:LTD}. Plasticity in layer 2/3 has also been found during the critical period of cat V1, and appears to precede modifications in the direct thalamic input to layer 4 during monocular deprivation experiments \cite{Trachtenberg:2001:RAP}). To what extent this involves changes in vertical rather than intralaminar horizontal connections has not yet been determined.

Another assumption of our model is that the CO blobs and OD columns emerge at roughly the same time, with the OD columns occurring slightly earlier. This is consistent with experimental findings, even if the precise temporal ordering is not yet clear. In macaque both CO blobs and OD columns emerge prenatally, so that at birth the pattern of OD columns and their spatial relationship with blobs is adult-like. In contrast, the cat's visual cortex is quite immature at birth. For example, supragranular layers of cat V1 differentiate postnatally and the blobs in these layers are normally first visible around 2 weeks of age (about one week after eye opening and one week before the critical period). This is approximately coincident with the earliest observation of OD columns in cat \cite{Crair:2001:EOD}.

\subsection{Mathematical formulation of model}

\begin{figure}[htbp]
\begin{center}
\includegraphics[width=0.9\linewidth]{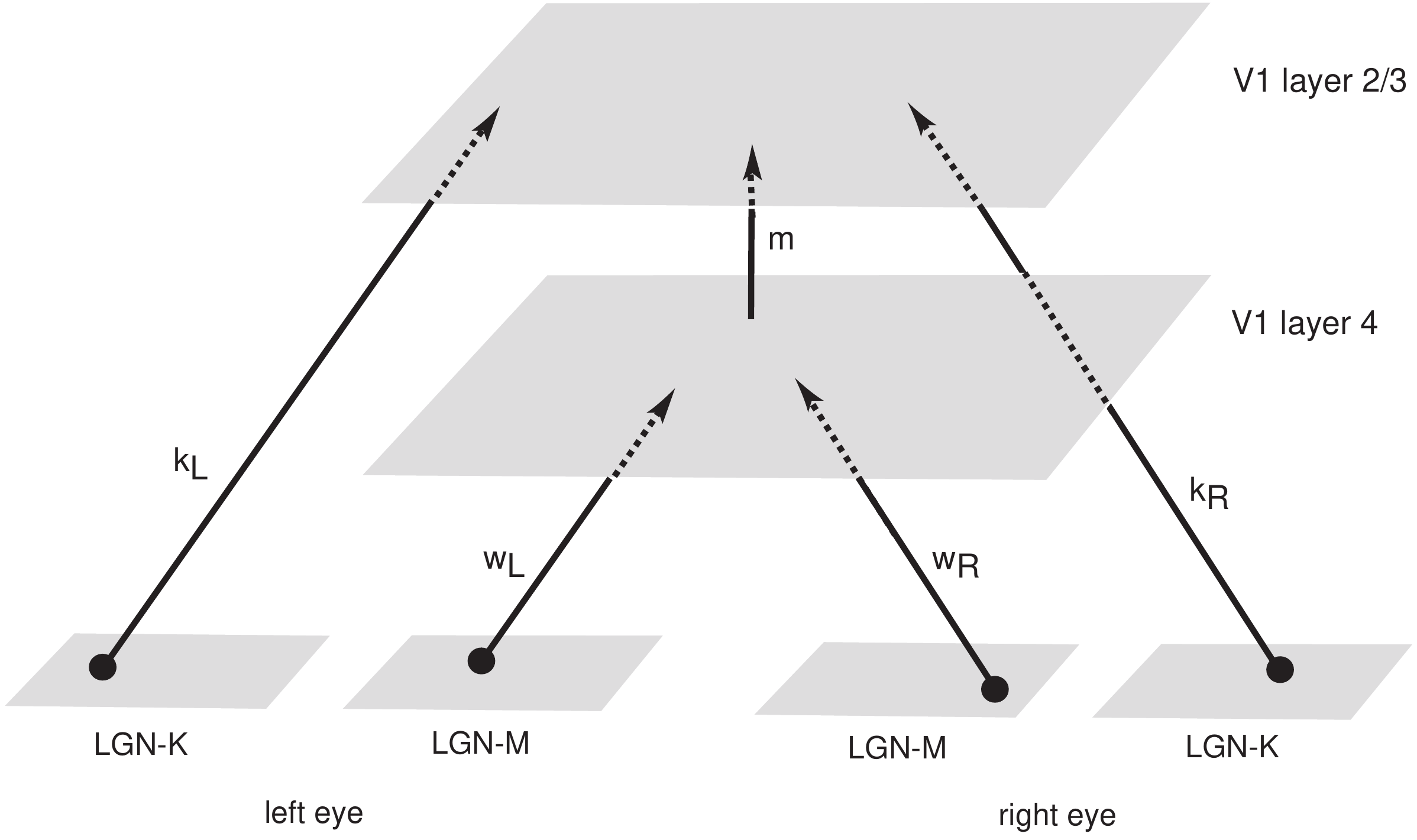}
\end{center}
\caption{\small 
Two--layer model of V1.}
\label{fig:model}
\end{figure}

In order to develop the mathematical formulation of our model we make two additional simplifications. First, we ignore the spread of
synaptic inputs between the LGN and layer 4C, and assume that there is a one--to--one retinotopic map between each LGN layer and each cortical layer. (This assumption will be relaxed later on, see section 3.2). Thus, after an appropriate
rescaling, we can represent each layer by $\Sigma \subset \R^2$ and uniquely label each point in an LGN layer by the corresponding point in cortex that it is connected to. Second, we consider only a single M-type
feedforward pathway from layer 4C to layer 2/3, which we assume innervates the interblobs in the mature cortex. Inclusion of a second M pathway together with a pair of P pathways as described in section 2.1 complicates the model without altering the basic mechanism identified in our study. 
A schematic diagram of the simplified model is shown in figure \ref{fig:model}. Let
$w_L(\rr)$ and
$w_R(\rr)$ denote the densities of left and right M-type LGN connections to a point $\rr=(x,y)$ of layer 4C in cortex at time $t$, and denote the corresponding K-type
densities from LGN to layer 2/3 by
$k_L(\rr)$ and
$k_R(\rr)$. Suppose that in each layer there are weak
cortical--cortical interactions between neurons at $\rr$ and $\rr'$ given by the synaptic density $J_i(\rr-\rr')$ where layers 4C and 2/3 are indexed by $i=1$ and $i=2$
respectively. There are also vertical feedforward connections from $M$-type neurons in layer 4C to layer 2/3 with synaptic density $m(\rr)$. 
Assuming a linear model for the cortical activity
$V_i(\rr,t)$ in the two layers, leads to the pair of coupled linear equations
\begin{equation}
\tau_v\frac{dV_1(\rr,t)}{dt}=-V_1(\rr,t)+\varepsilon \int_{\Sigma} J_1(\rr-\rr')V_1(\rr',t)d\rr'+w_L(\rr)I_L(\rr)+w_R(\rr)I_R(\rr) 
\label{V}
\end{equation}
and
\begin{equation}
\tau_v\frac{dV_2(\rr,t)}{dt}=-V_2(\rr,t)+\varepsilon \int_{\Sigma} J_2(\rr-\rr')V_2(\rr',t)d\rr'+k_L(\rr)\widehat{I}_L(\rr)+k_R(\rr)\widehat{I}_R(\rr) +m(\rr)V_1(\rr),
\label{hatV}
\end{equation}
where $I_{L,R}(\rr)$ and $\widehat{I}_{L,R}(\rr)$ denote the left and right inputs to the M and K layers of the LGN respectively. Note that although magnocellular and koniocellular
geniculate cells have different receptive field properties, we expect $I_L(\rr),\widehat{I}_{L}(\rr)$ to be strongly correlated, and similarly for the pair
$I_R(\rr),\widehat{I}_{R}(\rr)$. The constant coupling constant $\varepsilon$ determines the relative strength of the lateral connections.

Since development takes place on a much slower time--scale than the dynamics of cortical activity, we can take $V_i$ to be given by its steady--state value. However, calculating
the steady--state explicitly requires inverting the linear operator $\widehat{L}_i V_i(\rr)= V_i(\rr)-\varepsilon \int J_i(\rr-\rr')V_i(\rr')d\rr'$ for $i=1,2$. In the case of weak intracortical interactions, this inversion can be carried out by performing a perturbation expansion in $\varepsilon$. The first--order approximation is thus
\begin{equation}
V_1(\rr) =\int_{\Sigma} G_1(\rr-\rr') \left [w_L(\rr')I_L(\rr')+w_R(\rr')I_R(\rr') \right ]d\rr'
\label{Vss}
\end{equation}
and
\begin{equation}
{V}_2(\rr) =\int_{\Sigma} G_2(\rr-\rr') \left [k_L(\rr')\widehat{I}_L(\rr')+k_R(\rr')\widehat{I}_R(\rr') + m(\rr'){V}_1(\rr')\right ]d\rr'
\label{hatVss}
\end{equation}
with $G_i(\rr-\rr') \approx \delta (\rr-\rr') +\varepsilon J_i(\rr-\rr')$ and $\delta$ is the Dirac delta function. Note that we explicitly include the $\delta$-function component of the inverted operator $\widehat{L}_i ^{-1}$. Usually this term is ignored and $G_i$ is treated as a smooth weight function \cite{Miller:1989:ODC, Swindale:1996:DTV}.  As we show in the next section, the $\delta$--function can have a significant effect on the weight dynamics. Given the steady--state postsynaptic response to an input in the case of fixed weights, we now allow the feedforward weights to vary slowly in time according to a linear Hebb rule with subtractive normalization  \cite{Miller:1994:RCH}:

\begin{equation}
\tau_w \frac{dw_L}{dt} =\langle V_1I_L\rangle -\gamma({\mathbf w}) ,\quad \tau_w \frac{dw_R}{dt} =\langle V_1I_R\rangle -\gamma({\mathbf w}) 
\label{wHebb}
\end{equation}
and
\begin{eqnarray}
\label{kHebb}
\tau_k \frac{dk_L}{dt} &=&\langle V_2\widehat{I}_L\rangle -\eta({\mathbf k},m) ,\quad \tau_k \frac{dk_R}{dt} =\langle V_2\widehat{I}_R\rangle- \eta({\mathbf k},m), \\
\tau_m
\frac{dm}{dt} &=&\langle V_2V_1\rangle -\eta({\mathbf k},m), 
\label{mHebb}
\end{eqnarray}
where $\langle \ldots \rangle$ denotes averaging over the distribution of M-type inputs $I_{L,R}$ and K-type inputs  $\widehat{I}_{L,R}$. The decay terms $\gamma({\mathbf
w})$ and
$\eta({\mathbf k},m)$ enforce various conservation constraints whose specific form will be derived in section 3. Equations
(\ref{wHebb})--(\ref{mHebb}) are supplemented by additional constraints that ensure the weights remain positive and bounded. That is,
\begin{equation}
0 \leq w_{L,R} \leq W,\quad 0 \leq k_{L,R} \leq K,\quad 0\leq m \leq M .
\label{bound}
\end{equation}


\setcounter{equation}{0}
\section{Analysis of model}

\subsection{Development of OD columns in layer 4C}

Substitute equation (\ref{Vss}) into equation (\ref{wHebb}) and assume that the
M-type input correlations are of the form
\begin{equation}
\left (\begin{array}{cc} \langle I_L(\rr) I_L(\rr') \rangle  & \langle I_L(\rr) I_R(\rr') \rangle\\ \langle I_R(\rr) I_L(\rr') \rangle & \langle I_R(\rr) I_R(\rr')
\rangle
\label{corr}
\end{array}
\right ) = Q(\rr-\rr'){\mathbf C},\quad {\mathbf C}=\left (\begin{array}{cc} C_{LL}  & C_{RL}\\ 
                                                                                                                            C_{LR} & C_{RR}  \end{array} \right
).
\end{equation}
The matrix ${\mathbf C}$ determines the correlations between the two eyes.  We will take the eyes to be symmetric and to drive development equally.  That is, we take $C_{LL}=C_{RR}=C_S$ and 
$C_{LR}=C_{RL}=C_D$ so that $\mathbf C$  and is symmetric such that the diagonal terms are the same eye correlations $C_S$ and the
off--diagonal terms are the different eye correlations $C_D$. This leads to the vector equation
\begin{equation}
\tau_w\frac{d{\bf w}(\rr,t)}{dt}=\int_{\Sigma} H_1(\rr-\rr') {\mathbf C}{\mathbf w}(\rr',t)  d\rr' -\gamma({\mathbf w}) {\mathbf a},
\label{H1}
\end{equation}
where $H_1(\rr)=G_1(\rr)Q(\rr)$, ${\mathbf a} = (1,1)^T$ and ${\mathbf w} = (w_L,w_R)^T$. The subtractive term $\gamma(\omega) {\mathbf a}$ is chosen so that the total synaptic density $w_L(\rr)+w_R(\rr)$ is conserved at each point in cortex. Thus, we take
\begin{equation}
\label{gam}
\gamma({\mathbf w})=\mu\int_{\Sigma} H_1(\rr-\rr') \left [w_L(\rr)+w_R(\rr)\right ]d\rr' .
\end{equation}
Exploiting the fact that the input correlation matrix ${\mathbf C}$ has eigenvalues $\mu_{\pm}=C_S\pm C_D$ with corresponding eigenvectors ${\mathbf e}_{\pm} =(1,\pm 1)$, it is
straightforward to show that the vector equation (\ref{H1}) decomposes into the pair of decoupled equations
\begin{eqnarray}
\tau_w\frac{dw_{+}(\rr,t)}{dt}&=&(C_S+C_D -2\mu)\int_{\Sigma} H_1(\rr-\rr') w_{+}(\rr',t) d\rr' \\
\tau_w\frac{dw_{-}(\rr,t)}{dt}&=&(C_S-C_D)\int_{\Sigma} H_1(\rr-\rr')
w_{-}(\rr',t) d\rr'
\label{w}
\end{eqnarray}
with $w_{\pm}=w_L\pm w_R$.  Conservation of total synaptic density is thus achieved by setting $\mu =
\mu_+/2$ so that $dw_+(\rr,t)/dt =0$ for all $\rr$ and $t$.  For a more general choice of correlation matrix ${\mathbf C}$, the subtractive normalization term takes the form 
\begin{equation}
\label{subNorm}
 \gamma(\ww)= \frac{1}{ {\mathbf a} \cdot {\mathbf a}} \left ( {\mathbf a}\cdot \int_{\Sigma} H_1(\rr -\rr') \C \ww(\rr',t) d\rr' \right ).
\end{equation}
A higher dimensional version of equations (\ref{w}) and (\ref{subNorm}), involving a three--component weight vector, will be derived in section 3.2 when we consider development of layer 2/3. An additional complicating factor in the higher--dimensional case is that one of the synaptic components may vanish at a given point in cortex, and remain at zero due to the imposed lower bound on all the of weights, see equation (\ref{bound}). The subtractive normalization term then has to be modified so that the total synaptic density is still locally conserved. This issue is further discussed in section 3.2.

\begin{figure}[htbp]
 \begin{center}
\includegraphics[width=0.9\linewidth]{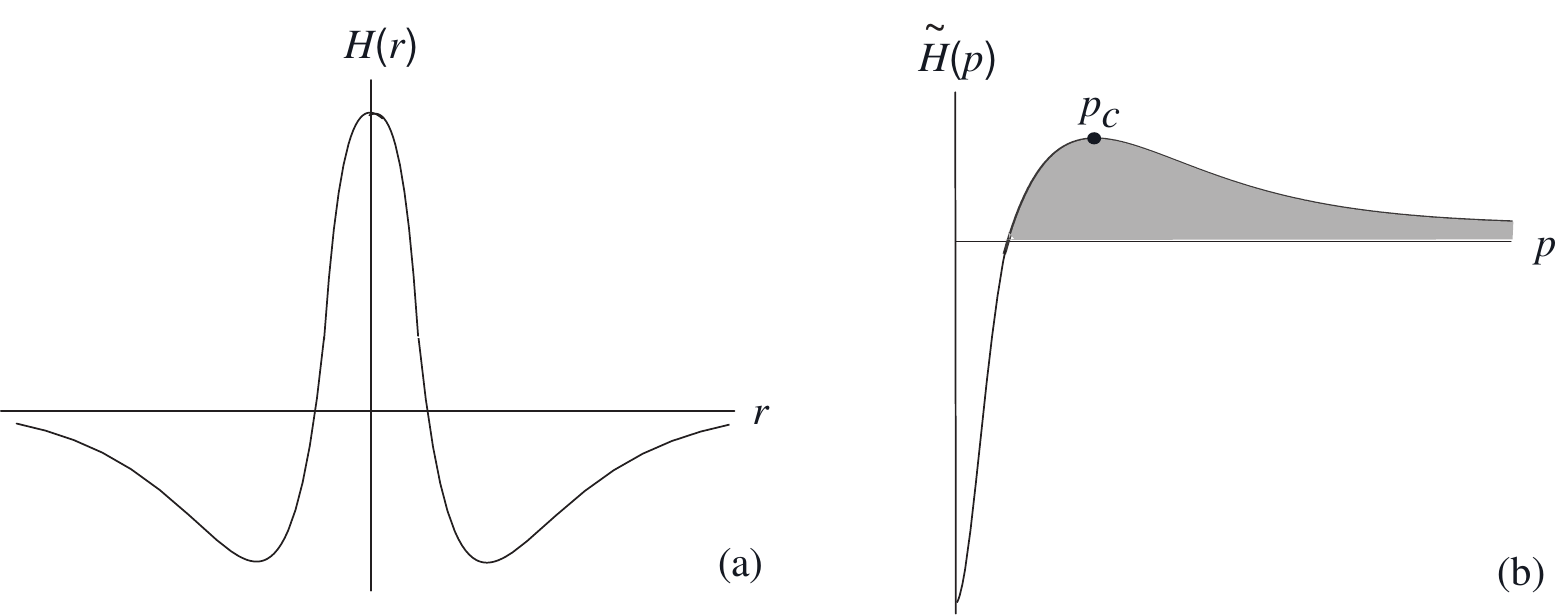}
    \caption{ \small (a) Difference--of--Gaussians
	      interaction function $H(r)$ displaying
	      short-range excitation and long-range inhibition. (b)
	      Fourier transform $\widetilde{H}(p)$ with maximum at $p=p_c$. The gray shaded region
	     denotes the semi--infinite band of unstable modes.}
\label{fig:DOG}
  \end{center}    
\end{figure}

The remaining equation for $dw_-/dt$ describes the growth of ocular dominance columns and can be analyzed using Fourier transforms. For the sake of illustration, suppose that $\Sigma$ is the unbounded two--dimensional plane. Then
\begin{equation}
w_{-}(\rr,t)=\int_{\R^2} A_p\e^{i{\mathbf p}\cdot \rr}\e^{\lambda(p) t/\tau_w}d\pp,
\label{w-}
\end{equation}
where $p=|{\mathbf p}| $ and $\lambda(p)= (C_S - C_D)\widetilde{H}_1(p)$ with
\begin{equation}
\widetilde{H}_1(p)=\int_{\Sigma} \e^{-i{\mathbf p}\cdot \rr}H_1(\rr)d\rr .
\end{equation}
The complex amplitude $A_p$ is determined by the initial
conditions:
\begin{equation}
A_p=\int_{\Sigma} \e^{-i{\mathbf p}\cdot \rr}w_-(\rr,0)d\rr .
\end{equation}
We expect correlations between the same eye to be greater than correlations between opposite eyes, that is, $C_S > C_D \geq 0$. Suppose that the Fourier transform
$\widetilde{H}_1(p)$, which is a real symmetric function of the wavenumber $p$, has a unique positive maximum at
$p=p_c$. 
The standard way of implementing this condition is to take $H_1({\mathbf r})=H(r)$ where $H$ consists of short--range excitation and long--range inhibition, as realized by either a
difference--of--Gaussians or a difference--of exponentials. The former takes the form
\begin{equation}
H(r) = A \left ( \e^{-|r^2| /2\sigma_E^2 }-\beta \e^{-{|r^2|}/{2\sigma_I^2} } \right ),\quad \sigma_E < \sigma_I,\quad 0 < \beta < 1,
\label{DOG}
\end{equation}
with $\sigma_E$ and $\sigma_I$ determining the range of excitation and inhibition respectively. An example of $H(r)$ and its transform $\widetilde{H}(p)$ is shown
in figure \ref{fig:DOG}. It can be seen that there is a
semi--infinite band of wavenumbers for which $\widetilde{H}(p) > 0$. It follows from equation (\ref{w-}) that there is an exponential growth of $w_-(\rr,t)$, with the fastest
growing mode consisting of a spatially periodic pattern of critical wavelength $2\pi/p_c$. The ultimate growth of the modes is restricted by the constraints $0 \leq
w_{L,R} \leq W$. Numerically one finds alternating bands of left/right eye ocular dominance columns whose mean spacing is determined by the critical wavelength (see also section
4). Layer 4C can thus be partitioned into left and right OD regions $\Sigma_{L,R}$ such that
$\ds \lim_{t\rightarrow \infty} {\mathbf w}(\rr,t)= {\mathbf W}(\rr)$ with ${\mathbf W}(\rr)=(W,0)^T$ for all $\rr \in \Sigma_L$ and ${\mathbf W}(\rr)=(0,W)^T$  for all $\rr \in \Sigma_R$. A similar analysis holds when $\Sigma$ is a bounded domain with the integral in equation (\ref{w-}) replaced by a discrete sum.

\subsubsection{Inclusion of $\delta$ component}

\begin{figure}[htbp]
\begin{center}
\includegraphics[width=0.9\linewidth]{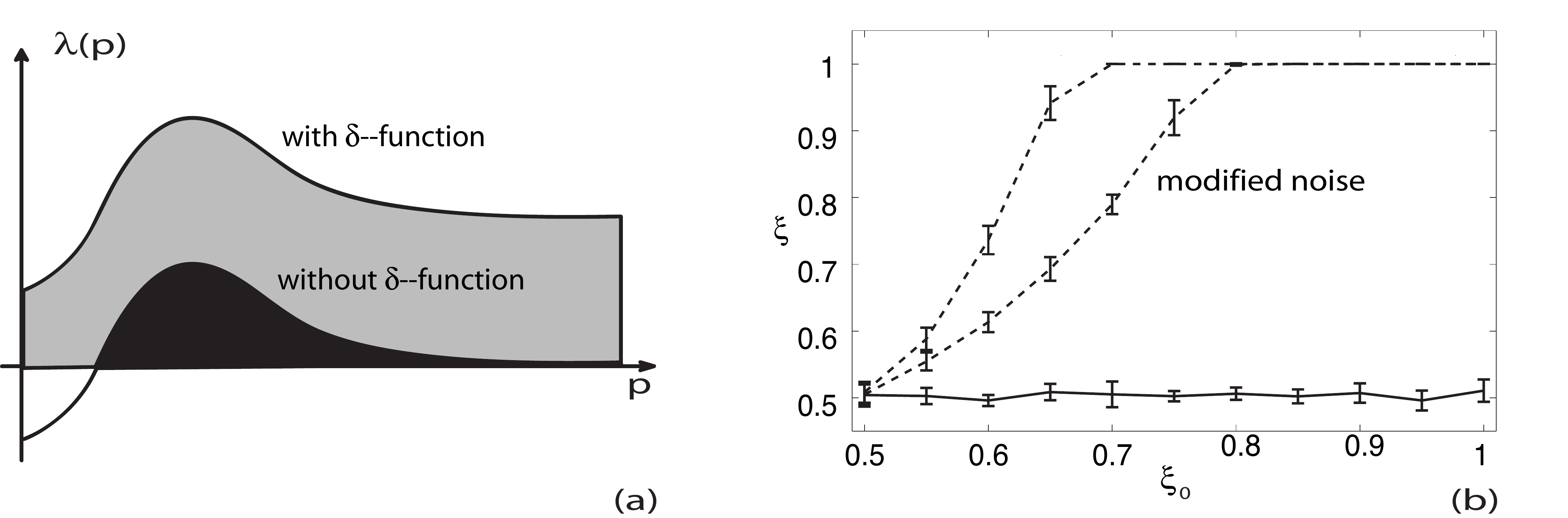}
\end{center}
\vspace{-5ex}
\caption{\small (a) Plot of $\widetilde{H}_1(p))$ as a function of wavenumber $p$ showing the upward shift induced by the inclusion of the Dirac delta function in the effective lateral interaction function $H_1({\mathbf r})$. Such a shift destabilizes the zero wavelength mode. (b) Ratio of resulting total left eye input to total input against the corresponding ratio for the initial state. We compare the resulting patterns for the cases where the $\delta$-component of $H_1$ is included or excluded.  Additionally, we consider the effects of filtered white noise when the $\delta$-component is included.  Starting from a spatially uniform state $w_L (r,0)= w_L^0$, $w_R(r,0)=1-w_L^0$ with $0\leq w_L^0 \leq 1$, we numerically solve equations (\ref{H1}) and (\ref{gam}) to obtain a steady--state OD pattern in layer 4C. The fraction $\xi$ of the pattern that is innervated by left eye inputs is averaged over 10 trials for fixed initial fraction $\xi_0$, and  then plotted as a function of $\xi_0$. Standard deviations are indicated by vertical bars. The dashed lines correspond to the result with the inclusion of the $\delta$-function component in $H_1$, whereas the solid line corresponds to the case without. Here $\varepsilon = 0.1$ and the parameters of the weight function (\ref{fig:DOG}) are 
$A=30,$  $\beta = 0.5$, $\sigma_E=0.4,$ and  $\sigma_I=0.62$.  
The modified noise term is taken to be of the form (\ref{eq:modNoise}), with a noise amplitude of 0.6 and a range of wave numbers between $p_{min}=1.5$ and $p_{max}=8.0$.  }
\label{fig:deltaStuff}
\end{figure}

One subtle aspect of the above analysis that is often overlooked concerns the inversion of the linear operator $\widehat{L}_1$, under the assumption that the lateral
interactions within layer 4C are relatively weak compared to the feedforward afferents from the LGN. In particular, $G_1(\rr)= \delta(\rr)+\varepsilon J_1(\rr)$ so that $H_1(\rr) =
Q(0)\delta(\rr) + \varepsilon Q(\rr)J_1(\rr)$. The contribution from the term $Q(0)\delta(\rr)$ is usually ignored and $Q(\rr)J_1(\rr)$ is identified with the Mexican hat function
$H(r)$. At first sight such a simplification does not appear to be significant, since the Fourier transform of the term $Q(0)\delta(\rr)$ simply introduces a
$p$--independent shift in $\widetilde{H}(p)$ so that the critical wavenumber $p_c$ remains the same, see figure \ref{fig:deltaStuff}(a).  However, such a shift destabilizes the uniform mode $p=0$, which then amplifies the homogeneous component of the initial state. The amplitude $A_0$ of this component will depend upon the initial balance between left and right eye inputs, since
$A_0=\int_{\Sigma}\{w_L(\rr,0)-w_R(\rr,0)\}d\rr$. We define a measure, $\xi$, for the total fraction of left eye input innervating layer 4C as
\begin{equation}
\ds   \xi = \frac{\int _{\Sigma} w_L(\rr)d\rr}{\int_{\Sigma} w_L(\rr)+w_R(\rr)d\rr }.
\end{equation}
The measure $\xi$ varies from 0 to 1, where 0 corresponds to total right eye dominance, 1 is total left eye dominance, and 1/2 is a balanced L/R state. A comparison of the dynamics with and without the $\delta$-function term is illustrated in figure \ref{fig:deltaStuff}(b). It can be seen that if the $\delta$--function contribution to $H_1$ is included, then the fraction of the resulting ocular dominance column pattern that is innervated by each eye is sensitive to the initial conditions. That is, if one eye starts out as dominant, then this dominance persists during the development of the OD columns. Typically, in numerical simulations of OD development, the initial state is taken to be approximately binocular so that this issue does not arise. However, when we consider the development of layer 2/3, the presence of $\delta$--function terms in the inverse operators $\widehat{L}_i^{-1}$ become significant  (see section 3.2).

In the case of OD development in layer 4C, the dependence on the initial balance between left and right eye inputs can be mitigated by adding a spatially varying component to a uniform initial state. For example, consider a spatially random perturbation $\xi(\rr)$ of each component of the initial state that consists of band--pass filtered white noise:
\begin{equation}
\label{eq:modNoise}
  \xi(\rr) ={\mathcal A}_{\xi} \int _{| \pp |=p_{min} }^{ p_{max} } \xi_{\pp} \cos(\pp \cdot \rr)d\pp,
\end{equation} 
where $p_{min}$ and $p_{max}$ determine the range of modes contained in the noise spectrum, ${\mathcal A}_{\xi}$ is the noise amplitude, and the $\xi_{\pp}$ are chosen from a uniform distribution on the interval [-1,1]. We consider band--passed rather than white noise in order to exclude additional contributions to the amplitude $A_0$. The presence of the noise term $\xi(\rr)$ induces a shift to the right of the sensitivity plot shown in figure \ref{fig:deltaStuff}(b). In order to understand this result, let us return to the Fourier decomposition (\ref{w-}) of $w_-(\rr,t)$. The growth of the OD pattern can be viewed as a race between all modes $p$ for which $\lambda(p) > 0$. Classically, one considers the mode that corresponds to the critical wavenumber $p_c$, i.e., the ``fastest'' growing mode, to dominate the final pattern that emerges once the saturating nonlinearities of equation (\ref{bound}) have been imposed.  With the inclusion of the $\delta$--component, the zeroth mode has also entered the race so that an approximately uniform initial state means that $|A_0|\gg |A_p|$ for all $p \neq 0$ which, in effect, gives the zeroth mode a large head start in the race. Thus, the resulting pattern becomes dependent on $A_0$ and hence 
the initial balance between left and right eye inputs. Let $T$ and $\widehat{T}$ represent the times for the critical mode $p_c$ and the zeroth mode, respectively, to reach the upper bound $W$. In order that segregation of the L and R visual pathways occurs, the parameter regime must be such that $T< \widehat{T}$. That is
\begin{equation}
  \frac{\tau_w}{\lambda(p_c)}\ln\left( \frac{W}{|A_{p_c}|} \right ) < \frac{\tau_w}{\lambda(0)}\ln\left( \frac{W}{|A_0|} \right ) .
\end{equation}
This inequality is satisfied provided that
\begin{equation}
\label{init_Ineq}
   \left ( \frac{|A_0|}{|A_{p_c}|} \right ) <  \left ( \frac{W}{|A_0|} \right ) ^ {\alpha_c},
\end{equation}
where $\alpha_c = \frac{\lambda(p_c)-\lambda(0)}{\lambda(0)}$. Note that if $|A_{p_c}|\ll |A_0|$, then the zeroth mode dominates and the cortex would develop into a wholly left or right eye driven state. Inclusion of the noise term $\xi(\rr)$ in the initial state increases the ratio $|A_{p_c}|/|A_0|$ so that the inequality (\ref{init_Ineq}) is satisfied.

\subsection{Development of OD columns and CO blobs in layer 2/3}

As the ocular dominance columns form in layer 4C, they influence the development of the feedforward afferents $m(\rr)$ connecting layer 4C to layer 2/3 as well as the
koniocellular afferents $k_{L,R}(\rr)$  projecting from the LGN to layer 2/3. Motivated by experimental data in the cat \cite{Katz:1992:DLC}, we will assume that the OD columns in layer 4C form before the OD columns and CO blobs in layer 2/3 so that we can take ${\mathbf w}(\rr,t)\rightarrow {\mathbf W}(\rr)$ in equations (\ref{kHebb}) and (\ref{mHebb}). (Note that even if they were to develop simultaneously, the analysis would still be valid to a lowest order approximation, since there is no direct feedback from layers 2/3 to layer 4C). Following along similar lines to the analysis of equation (\ref{wHebb}), we introduce the input correlation matrices
$\widehat{\mathbf C}$ and
${\mathbf B}$ according to  
\begin{equation}
\langle \widehat{I}_{i}(\rr) \widehat{I}_j(\rr') \rangle = Q(\rr-\rr')\widehat{\mathbf C}_{ij}, \quad \langle {I}_{i}(\rr) \widehat{I}_j(\rr') \rangle =
Q(\rr-\rr')B_{ij}
\end{equation}
for $i = R,L$ and $j=R,L$ with
\begin{equation}
\widehat{\mathbf C}=\left (\begin{array}{cc} \widehat{C}_S  & \widehat{C}_D\\ \widehat{C}_D & \widehat{C}_S  \end{array} \right
),\quad {\mathbf B}=\left (\begin{array}{cc} B_S  & B_D\\ B_D & B_S  \end{array} \right).
\end{equation}
For simplicity, we assume that the $M/M$, $K/K$ and $K/M$ input correlations all have the same spatial dependence as determined by the function $Q(\rr)$. 
Evaluating equations (\ref{Vss}) and (\ref{hatVss}) gives 
\begin{eqnarray}
\label{VIa} 
\langle V_2(\rr,t) \widehat{\mathbf I}(\rr)\rangle &=& \int G_2(\rr-\rr') \widehat{\mathbf C} {\mathbf k}(\rr',t) Q(\rr'-\rr) d\rr'\\
&+&\int G_2(\rr-\rr') m(\rr',t)G_1(\rr'-\rr''){\mathbf B} {\mathbf W}(\rr'') Q(\rr''-\rr) d\rr''d\rr' \nonumber
\end{eqnarray}
and
\begin{eqnarray}
\label{VVa}
 \langle V_2(\rr,t) V_1(\rr,t)\rangle  = & \int G_2(\rr-\rr') {\mathbf k}(\rr',t)\cdot {\mathbf B} {\mathbf W}(\rr'') Q(\rr'-\rr'') G_1(\rr''-\rr)d\rr''d\rr'  \nonumber \\ 
 &+\int G_2(\rr-\rr') m(\rr',t)G_1(\rr'-\rr'') {\mathbf  W}(\rr'')\cdot {\mathbf C} {\mathbf  W}(\rr''')  \dots  \nonumber \\
  &  Q(\rr''-\rr''') G_1(\rr'''-\rr)d\rr'''d\rr''d\rr'.
\end{eqnarray}

It is clear from equations (\ref{VIa}) and(\ref{VVa}) that when considering correlation--based Hebbian learning in a multi--layer network, the resulting dynamical equations for the weights rapidly become complicated due to the presence of multiple convolutions. This makes it difficult to gain insights into the basic mechanisms underlying feature map formation in the higher layers, and also makes numerical simulations more difficult. However, considerable simplification can be obtained by carrying out a perturbation expansion to first order in the coupling parameter $\varepsilon$. Such an approximation is already implicit in the inversion of the linear operators $\widehat{L}_i$. In order to proceed we will need to include the Dirac delta function in $\widehat{L}_i^{-1}$. Indeed, considerable insight into the underlying linear algebraic structure of the Hebbian dynamics can be
obtained by setting $\varepsilon =0$ in equations (\ref{VIa}) and (\ref{VVa}) so that $G_i(\rr-\rr')\rightarrow \delta(\rr-\rr')$ . This corresponds to ignoring the effects of lateral interactions; such interactions will subsequently be incorporated by carrying out a first--order perturbation expansion in $\varepsilon$.  

\subsubsection{${\mathcal O}(1)$ analysis} 

The zeroth--order contributions to the averages in equations (\ref{VIa}) and (\ref{VVa}) are given by
\begin{eqnarray}
\label{VI}
\langle V_2(\rr,t) \widehat{\mathbf I}(\rr)\rangle_0 &=&  Q(0)\left [\widehat{\mathbf C} {\mathbf k}(\rr,t) +m(\rr,t){\mathbf B} {\mathbf  W}(\rr) \right ]
\end{eqnarray}
and
\begin{eqnarray}
\label{VV}
 \langle V_2(\rr,t) V_1(\rr,t)\rangle_0 = Q(0)\left [ {\mathbf k}(\rr,t)\cdot {\mathbf B} {\mathbf  W}(\rr)  +
 m(\rr,t){\mathbf  W }(\rr)\cdot {\mathbf C} {\mathbf W}(\rr) \right ].
\end{eqnarray}
Substituting equations (\ref{VI}) and (\ref{VV}) into (\ref{kHebb}) and (\ref{mHebb}) then gives (after rescaling such that $W=1$, $Q(0)=1$)
\begin{equation}
\tau_k \frac{dk_L}{dt}=\widehat{C}_S k_L(\rr,t) + \widehat{C}_D k_R(\rr,t) + [ B_S \chi_L(\rr)+B_D\chi_R(\rr)] m(\rr,t) -\eta_0({\mathbf k},m)
\label{kL}
\end{equation}
\begin{equation}
\tau_k \frac{dk_R}{dt} = \widehat{C}_S k_R(\rr,t) + \widehat{C}_D k_L(\rr,t) +  [ B_D \chi_L(\rr)+B_S\chi_R(\rr)] m(\rr,t) -\eta_0({\mathbf k},m)
\label{kR}
\end{equation}
\begin{eqnarray}   
\tau_m \frac{dm}{dt} &=&  \( {C}_{S} \biggr [ \chi_L^2(\rr) +\chi_R^2(\rr) \biggr ]+2 \chi_L \chi_R  C_D \biggr )  m(\rr,t) + \left (B_S\chi_L(\rr)+B_D\chi_R(\rr) \right ) k_L(\rr,t)\nonumber \\
&& + \left (B_D\chi_L(\rr)+B_S \chi_R(\rr)\right ) k_R(\rr,t) -\eta_0({\mathbf k},m),
\label{m}
\end{eqnarray}
where $\eta_0$ is the subtractive normalization term when $\varepsilon=0$, and $(\chi_L,\chi_R)=(1,0)$ or $(0,1)$ depending upon whether $\rr \in \Sigma_L$ or $\rr \in \Sigma_R$, respectively.

The binary-valued functions $\chi_{L,R}(\rr)$ arise from the simplification that there is a one--to--one vertical interlaminar projection from layer 4C to layer 2/3. We now relax this assumption by allowing the vertical connections to sample over a localized region of layer 4C. The sampling region reflects the spatial extent of the dendritic field of the forward projecting neurons, which we assume is a fixed arbor function. We can incorporate the effect of a dendritic field into the above model by taking the functions $\chi_{L,R}(\rr)$ to represent the degree of left and right eye drive from layer 4C projected to point $\rr$ in layer 2/3 such that $0 \leq \chi_{L,R}(\rr) \leq 1$.  More specifically, introducing the normalized Gaussian $J_f(\rr)$ with space constant $\sigma_f$ and setting \\ $W=1$, we consider
\begin{equation}
 \chi_{L,R}(\rr) = \Phi   \biggr ( \int J_f(\rr-\rr') W_{L,R} (\rr')d\rr' \biggr )
   \label{chiJ}
\end{equation}
Under this formulation, the synaptic input to a cortical location $\rr$ due to the left and right eye streams is sampled from the footprint and is subject to a transfer function $\Phi$, where we take $\Phi(\nu)=\sqrt{\nu}$.  With this choice of $\Phi$, $\chi_L^2(\rr) +\chi_R^2(\rr)=1$ for all $\rr$.  Naturally the smooth variation in $\chi_{L,R}$ sampled from the above layer naturally results in the emergence of binocular regions along the OD borders in layer 2/3 (discussed below).  As there is likely a similar integration of left and right eye drive occurring in layer 4C due to input from the LGN and accounting for cortical-thalamic feedback (reviewed in \cite{Sherman:2017}), we use this formulation $\chi_{L,R}(\rr)$ in equations \ref{kL}, \ref{kR}, \ref{m}.  

In addition, we track the balance of left versus right eye drive (and vice-versa) at a position $\rr$ in the upper laminae by introducing the measures, 
\begin{equation}
   \Upsilon_{L}(\rr) = \frac{\chi_L(\rr)} {\chi_L(\rr) +\chi_R(\rr)}   \text{ and }   \Upsilon_{R}(\rr) = \frac{\chi_R(\rr)} {\chi_L(\rr) +\chi_R(\rr)} 
\end{equation}
Near the center of a left-eye OD column $\Upsilon_L(\rr) \approx 1$ and  $\Upsilon_R \approx 0$, near the center of a right-eye OD column $\Upsilon_L(\rr) \approx 0$ and  $\Upsilon_R \approx 1$, and near the OD border $\Upsilon_L(\rr) \approx \Upsilon_R \approx 1/2$.

In the case of primates, the development of OD columns and CO blobs occurs prenatally so that the Hebbian process is
driven by spontaneous activity in the left and right eyes. Therefore, we expect opposite eye correlations to be small and we can set $C_D=\widehat{C}_D =B_D=0$ for simplicity.  As a further simplification we also set $\widehat{C}_S = C_S > B_S$, although our results generalize to the case $C_S \neq \widehat{C}_S$.  We then rewrite equations (\ref{kL})--(\ref{m}) in the matrix form
\begin{eqnarray*}
\frac{d}{dt}\left (\begin{array}{c} k_L\\k_R \\ m \end{array}\right ) = \left ( \begin{array}{ccc}
                                       \widehat{C}_S & 0 & B_S \chi_L \\
                                        0 & \widehat{C}_S & B_S \chi_R \\                                         
                                         B_S \chi_L  &   B_S \chi_R  & C_S \end{array} \right )
                                         \left (\begin{array}{c} k_L\\k_R \\ m \end{array}\right )-\eta_0({\mathbf k},m)\left (\begin{array}{c} 1\\ 1 \\ 1 \end{array}\right ),
\label{dk}
\end{eqnarray*}
where we have set $\tau_k=\tau_m =1$ for convenience and dropped the explicit dependence on $\rr$. Following along analogous lines to the analysis of OD column formation in layer 4C, we choose the subtractive normalization term $\eta_0$ in order to conserve the total synaptic density $k_L+k_R +m$ at each point in layer 2/3. However, there is an additional subtlety that arises when considering three rather than two weight components: the normalization term must
be redefined in the event of one component vanishing. As we show below, this occurs in the center of OD columns that are inherited from layer 4C, where either $k_R$ or $k_L$ vanishes and the remaining component subsequently competes with the $m$ input. In order to conserve the total synaptic density throughout the developmental process, we define the subtractive normalization in a dynamic way:
\begin{eqnarray*}
\label{eq:etas}
 \eta_0 & = &\theta(k_L)\theta(k_R) \eta_0^1 + [ 1-\theta(k_L)\theta(k_R) ]\eta_0^2 ,  \\
        \eta_0^1 &= & \frac{1}{3}  ( \ C_S \left [ k_L+k_R+m \right )]
                              + B_S \left [ \chi_L \left ( k_L+m \right ) +\chi_R \left ( k_R+m \right )
                 \right ]  \ ) , \\
     \eta_0^2 & = &\frac{1}{2} (  \  C_S \left [ k_L+k_R+m \right )]   
                                 + B_S \left [ \theta(k_L) \chi_L \left ( k_L+m \right )
                                         +\theta(k_R) \chi_R  \left ( k_R+m \right ) \right ]   \  )  ,     
   \end{eqnarray*}
where $\theta$ is a Heaviside function. With  this choice of an adaptive normalization scheme, in effect, we are taking the standard choice for the subtractive normalization term for the remaining components after one component vanishes, see equation (\ref{subNorm}).

We  illustrate the behavior of the solutions in two particular cases: $\Upsilon_L(\rr) = 1$ and 1/2. Here $\Upsilon_L(\rr) = 1$ corresponds to a point in layer 2/3 that is above the center of a left-eye OD column in layer 4C and $\Upsilon_L(\rr)=1/2$ corresponds to a point above an OD border.  We take the initial conditions to be $k_L(t=0)=k_R(t=0)=k_0$, $m(t=0)=m_0$ with $k_0=m_0$.

\begin{figure}[htbp]
\begin{center}
      \includegraphics[width=\linewidth]{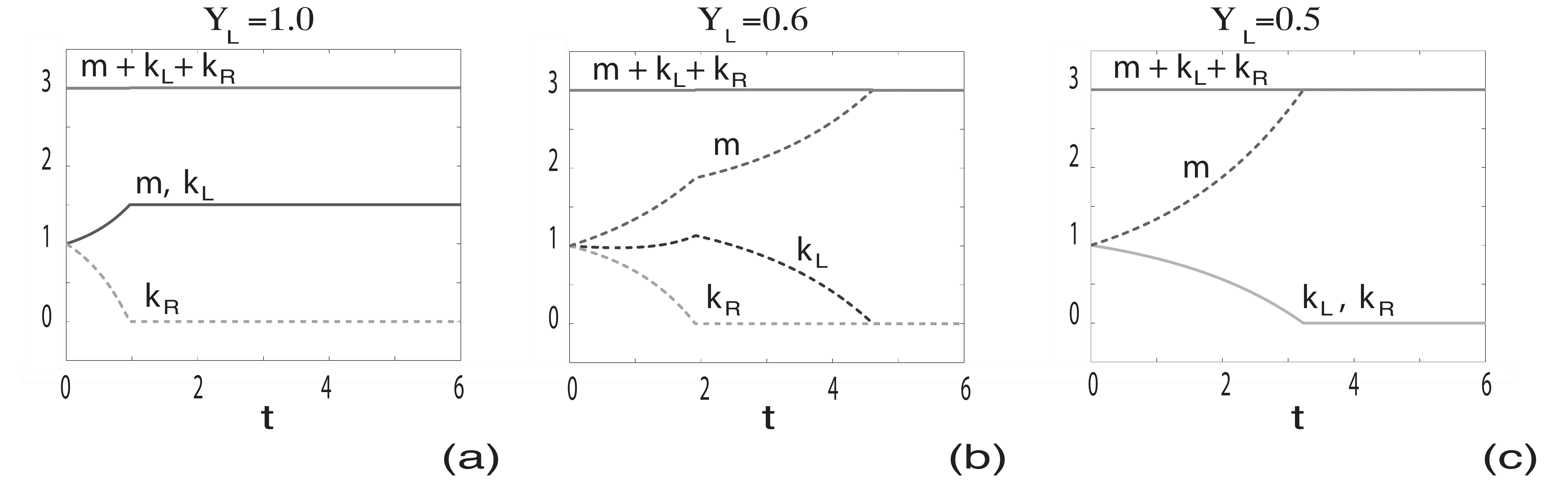}
 \vspace{-0.5cm}
\caption { \small Solution of ${\mathcal O}(1)$ dynamics for various values of $\chi$ with the initial values of $k_L$, $k_R$, and $m$ all equal to 1.  The same eye correlation coefficients are $C_S=\widehat{C}_S=1.0$ and $B_S=0.8$, and all alternate eye correlations are taken to be 0.  (a) Center of a left-eye column ($\Upsilon  =1$). (b) Close to an OD boundary within a left-eye column ($\chi_L=0.6$). (c) At an OD border ($\chi_L=0.5$).}
\label{fig:orderOne}
\end{center}
 \end{figure}

\noindent \textit{Case I: Center of left-eye column} ($\Upsilon_L(\rr)=1:  \ \chi_L=1, \ \chi_R=0$) Under the change of variables $k_{\pm} = k_L{\pm}m$, we have
\begin{eqnarray}
 \frac{dk_+}{dt}&=&  \frac{1}{3}[(C_S+B_S)k_+  - 2 C_S k_R],
\label{OD1}
\\ \frac{dk_-}{dt}&= &  (C_S-B_S)  k_- ,
\label{OD2}
 \\
 \frac{dk_R}{dt}&= & \frac{1}{3}[-(C_S+B_S)k_+  +  2 C_S k_R].
\label{OD3}
\end{eqnarray}
It can be seen that the $k_-$ dynamics decouples, whereas the components $k_+$ and $k_R$ compete
 against one another.  We find that for a wide range of parameter values the right-eye population ($k_R$) vanishes leaving the left-eye population ($k_+$) intact.  The adaptive normalization scheme then leads to the new dynamics
\begin{eqnarray}
 \frac{dk_+}{dt}&=&  0
\label{OD21}
\\ \frac{dk_-}{dt}&= & (C_S-B_S)  k_- ,
\label{OD22}
\end{eqnarray}
in which the remaining left-eye K and M inputs now compete with one another.  We simulate the dynamics for a representative choice of parameters and display the results in figure \ref{fig:orderOne}(a).

\noindent \textit{Case II: OD border} ($\Upsilon_L(\rr) =1/2: \ \chi_L=\chi_R= \chi $) Given the constraint that $k_L+k_R+m$ is locally conserved, we can reduce the dynamics at the given point $\rr$ to the following pair of equations
\begin{equation}
   \frac{d}{dt} \left ( \begin{array}{c}  k_{tot}  \\   m   \end{array} \right )= \frac{1}{3} \left ( \begin{array}{cc} 
   									 (C_S-2 \chi B_S) &  (2 \chi B_S-2C_S) \\
									 (2 \chi B_S-C_S)  &(2C_S-2 \chi B_S) \end{array} \right )
							\left ( \begin{array}{c} 
   									 k_{tot}  \\   m   \end{array} \right )
\end{equation}
where $ k_{tot} =  k_L+k_R$. The associated matrix is degenerate and has the following non-zero eigenvalue
\begin{equation}
   \lambda =  \frac{1}{3}( 3C_S-4 \chi B_S) 
\end{equation}
Provided $3C_S-4 \chi B_S > 0$, the K and M pathways will compete as the associated eigenvector is
\begin{equation}
   {\mathbf v}= \left ( \begin{array}{c}  -1  \\  1   \end{array} \right )    
\end{equation}
For the given initial conditions, we find that $k_L,k_R\rightarrow 0$ at the OD border leaving a purely interlaminar, binocular magnocellular drive to the cortical point $\rr$, see figure \ref{fig:orderOne}(c). Similar behavior occurs in a region surrounding the OD border where $\Upsilon_L(\rr) \approx0.5$, see figure \ref{fig:orderOne}(b).

\bigskip

\noindent The above examples show that, given an approximately balanced uniform initial state, layer 2/3 rapidly inherits the OD columns of layer 4, with $k_R(\rr)= 0$ when $\rr \in \Sigma_L$ and $k_L(\rr) = 0$ when $\rr \in \Sigma_R$. Moreover, with the inclusion of the dendritic sampling from layer 4C, the koniocellular weights $k_{L,R}$ both decay to zero at the OD borders where $\Upsilon_L, \Upsilon_R \approx 0.5$ (see figure \ref{fig:orderOne}(c)), ultimately sequestering the CO blobs away from the OD borders. When we include spatial interactions, the competition within an OD column between the $K$--type and $M$--type inputs will result in the segregation of the K and M pathways, i.e., the formation of the blob/interblob areas.

\subsubsection{${\mathcal O}(\varepsilon)$ analysis} 

In order to incorporate the leading order effects of the lateral interactions, we now determine the ${\mathcal
O}(\varepsilon)$ contributions to equations (\ref{VIa}) and (\ref{VVa}):
\begin{eqnarray*}
\label{VI1}
\langle V_2(\rr,t) \widehat{\mathbf I}(\rr)\rangle_1 = K_2*\left [\widehat{\mathbf C} {\mathbf k}(\rr,t)+m(\rr,t){\mathbf B} {\mathbf W}(\rr)\right ]
+m(\rr)K_1*\left [{\mathbf B} {\mathbf W}(\rr)\right ]
\end{eqnarray*}
and
\begin{eqnarray*}
\label{VV1}
\langle V_2(\rr,t) V_1(\rr,t)\rangle_1 &=&{\mathbf W}(\rr)\cdot K_2*\left [ {\mathbf B} {\mathbf k}(\rr,t)+m(\rr,t) {\mathbf C}{\mathbf W}(\rr)\right ]
\nonumber \\ && +{\mathbf k}(\rr,t)\cdot K_1*\left [{\mathbf B}{\mathbf W}(\rr)\right ]+2m(\rr,t){\mathbf W}(\rr)\cdot K_1*\left [{\mathbf C}{\mathbf W}(\rr)\right ].
\end{eqnarray*}
We have introduced the convolution operator $*$ defined according to
\begin{equation}
K_i*[f(\rr)] = \int K_i(\rr-\rr')f(\rr')d\rr'
\end{equation}
for any function $f(\rr)$ where $K_i(\rr)=J_i(\rr)Q(\rr)$ for $i=1,2$. As in the analysis of the zeroth order dynamics, we assume for simplicity that $B_D=\widehat{C}_D =0$ and $\widehat{C}_S=C_S$.
Substituting the approximations $\langle V_2 {\mathbf I}\rangle = \langle V_2 {\mathbf I}\rangle_0+\varepsilon \langle V_2 {\mathbf I}\rangle_1$ and $\langle V_2 V_1\rangle =
\langle V_2 V_1\rangle_0+\varepsilon \langle V_2 V_1\rangle_1$ into equations (\ref{kHebb}) and (\ref{mHebb}), and using equations (\ref{VI}), (\ref{VV}), (\ref{VI1}),
(\ref{VV1}) then gives 
\begin{eqnarray}
\frac{dk_L}{dt} &=& {C}_S k_L(\rr,t) +B_S m_L(\rr,t) -\eta({\mathbf k},m)\nonumber \\ 
&&+\varepsilon K_2*\left [{C}_S k_L(\rr,t)+B_Sm_L(\rr,t)\right ]+
\varepsilon B_Sm(\rr)\psi_L(\rr)
\label{kL2}
\end{eqnarray}
\begin{eqnarray}
\frac{dk_R}{dt} &=& {C}_S k_R(\rr,t) +B_S m_R(\rr,t)-\eta({\mathbf k},m) \nonumber \\ 
&& +\varepsilon K_2*\left [{C}_S
k_R(\rr,t)+B_S m_R(\rr,t)\right ]+\varepsilon B_Sm(\rr)\psi_R(\rr)
\label{kR2}
\end{eqnarray}
\begin{eqnarray}
 \frac{dm}{dt} &=& {C}_S m(\rr,t) + B_S\chi_L(\rr) k_L(\rr,t)+ B_S \chi_R(\rr)k_R(\rr,t)-\eta({\mathbf k},m)  \nonumber \\
&& +\varepsilon B_S\left (\chi_L(\rr) K_2*k_L(\rr,t)+\chi_R(\rr)K_2*k_R(\rr,t)\right )\nonumber \\
&&+\varepsilon C_S \left (\chi_L(\rr)K_2* [m_L(\rr,t)]+ 
\chi_R(\rr)K_2*[m_R(\rr,t)]\right )\nonumber \\
&&+\varepsilon B_S\left (k_L(\rr,t)\psi_L(\rr)+k_R(\rr,t)\psi_L(\rr)\right )\nonumber \\
&& +2\varepsilon  C_S \left ( m_L(\rr,t) \psi_L(\rr)+ m_R(\rr,t) \psi_R(\rr) \right ).
\label{m2}
\end{eqnarray}
Here we use the notation
\begin{equation}
   m_L(\rr,t) = \chi_L(\rr) m(\rr,t) \ \ \ \ \ \text{and} \ \ \ \ \  m_R(\rr,t) = \chi_R(\rr)m(\rr,t) 
\end{equation}
and set
\begin{equation}
   \psi_L(\rr) = K_1*\Upsilon_L(\rr) \ \ \ \ \ \text{and} \ \ \ \ \ \psi_R(\rr) = K_1*\Upsilon_R(\rr).
\end{equation}

As in the ${\mathcal O}(1)$ case, we have to consider an adaptive normalization scheme in order locally to conserve total synaptic density throughout the developmental process.  Thus $\eta$ is of the form
\begin{equation}
\eta  = \theta(k_L)\theta(k_R) [\eta_0^1+\varepsilon\eta_1^1] + [ 1-\theta(k_L)\theta(k_R) )][\eta_0^2 +\varepsilon\eta_1^2] ,
\end{equation}
where $\eta_i^j$ denotes the subtractive normalization term for the $i$th order contribution at the $j$th stage. The terms $\eta_1^j$ are chosen so as to conserve total synaptic density at ${\mathcal O}(\varepsilon)$. The ${\mathcal O}(1)$ analysis showed how the layer 2/3 system can rapidly
inherit the ocularity preference of layer 4C.  We will assume that the OD segregation is maintained at ${\mathcal O}(\varepsilon)$ on taking the subtractive normalization $\eta$ as defined above.  The preservation of the OD map in layer 2/3 will be confirmed numerically in section 4.
The ${\mathcal O}(1) $ analysis also showed how around an OD border the koniocellular inputs vanish leaving only the contribution from the $M$ pathway, whereas in the central region of an OD column there is competition between an $M$ input and  left or right-eye K input. This suggests that the main effect of the ${\mathcal O}(\varepsilon)$ lateral interactions is to spatially segregate the $M$ and $K$ inputs within the central region of an OD column. The occurrence of such a segregation, which is confirmed numerically in section 4, generates the CO blobs. To further illustrate this, consider a point $\rr$ at the center of a left-eye dominated region where $\chi_L\approx 1$ and $\chi_R\approx 0$. Equations (\ref{kL2})--(\ref{m2}) reduce to the simpler form
\begin{eqnarray}
\frac{dk_{+}^L(\rr,t)}{dt}&=& 0   \nonumber \\ \\
\frac{dk_{-}^L(\rr,t)}{dt}& \approx &(C_S-B_S)\int_{\Sigma_L} H_2(\rr-\rr')k_{-}^L(\rr',t) d\rr' ,
\label{kkL}
\end{eqnarray} 
where $k_{\pm}^{L}=k_L\pm  m$. A similar equation holds for $\rr$ at the center of a right--eye dominated region under the change $L \rightarrow R$. The fact that the convolution with respect to the interaction kernel $H_2$ in equation (\ref{kkL}) is restricted to same--eye ocular dominance strips ($\rr, \rr' \in \Sigma_L$) rather than the whole two--dimensional layer, suggests that the pattern formation process is quasi--one--dimensional. That is the center of each OD column can be idealized as a narrow stripe that is partitioned into alternating regions of $K$ and $M$ afferents, that is, blobs and interblobs. To a first approximation, the spacing of the pattern within a single stripe will be determined by the Fourier transform of $H_2$ along the stripe. Weak interactions between same--eye stripes will then determine the alignment of patterns across the cortex. This basic picture is consistent with the numerical simulations presented in section 4. As in layer 4C, we take the effective interaction function to be $H_2=\delta+\varepsilon H$ with $H$ given by the Mexican hat function (\ref{DOG}). Hence, as in the layer 4C model, the presence of the Dirac delta function means that the resulting CO blob pattern will be sensitive to the initial balance of $M$ and $K$ afferents. This is handled in our numerical simulations (see below) by adding to the uniform initial state a modified white noise term analogous to equation (\ref{eq:modNoise}).

\section{Numerical results}

\begin{figure}[htbp]
\begin{center}
\includegraphics[width=0.9\linewidth]{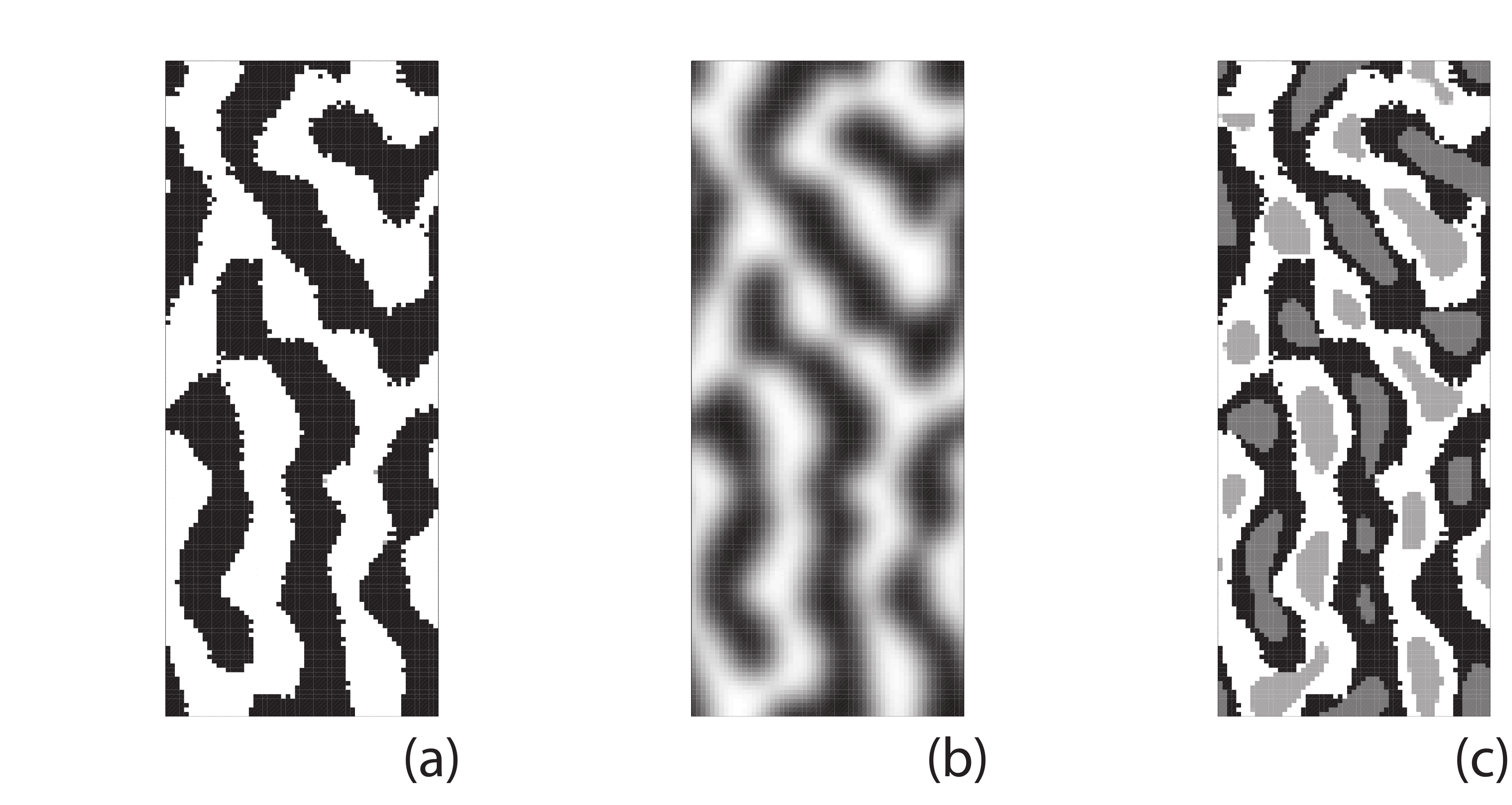}
\vspace{-0.5cm}
\caption{\small (a) OD pattern in layer 4C obtained by solving equations (\ref{H1}) and (\ref{gam}) starting from a homogeneous binocular state. Same parameters as in figure \ref{fig:deltaStuff}(b) with $w_L^{0}=0.5$.  (b) The projection of OD properties to layer 2/3 with a continuous variation in the OD properties.  In (a) and (b), white corresponds to left-eye M input, black corresponds to right-eye M input, and the grey areas represent a balance of left and right eye drive.  (c)  The resultant pattern in layer 2/3 displaying OD preference as in layer 4C as well as the segregation of the K and M pathways with the blob regions centered in the OD stripes.  In (c), white corresponds to left-eye dominated M input, black corresponds to right-eye dominated M input, and the light and dark grey areas are the left and right eye K inputs, respectively. OD pattern is obtained by solving equations (\ref{kL2})--(\ref{m2}) for $\psi_L,\psi_R=0$ starting from a homogeneous initial state with $k_L=k_R=9.7$ and $m=10.0$. The lateral interaction function $K_2$ is taken to be the Mexican hat function (\ref{fig:DOG}) with $A=8.0, \beta=1/2, \sigma_{E}=0.50, \sigma_{I}=0.78$, and $\varepsilon$ is taken to be 0.1.   The space constant of the Gaussian footprint in equation (\ref{chiJ}) is $\sigma_f = 0.4$.  The correlation parameters are $\widehat{C}_S=C_S=1.0$ and $B_S=0.4$.  The domain is 6 x 15 and was chosen to simulate a 2.3 $mm$ by 5.8 $mm$ region of cortex.  We use modified noise as given by equation  (\ref{eq:modNoise}) with $p_{min}=1.0$, $p_{max}=8.0$ and noise amplitude 0.5. }
\label{fig:finalPattern}
\end{center}
\end{figure}

\begin{figure}[htbp]
\begin{center}
\includegraphics[width=0.9\linewidth]{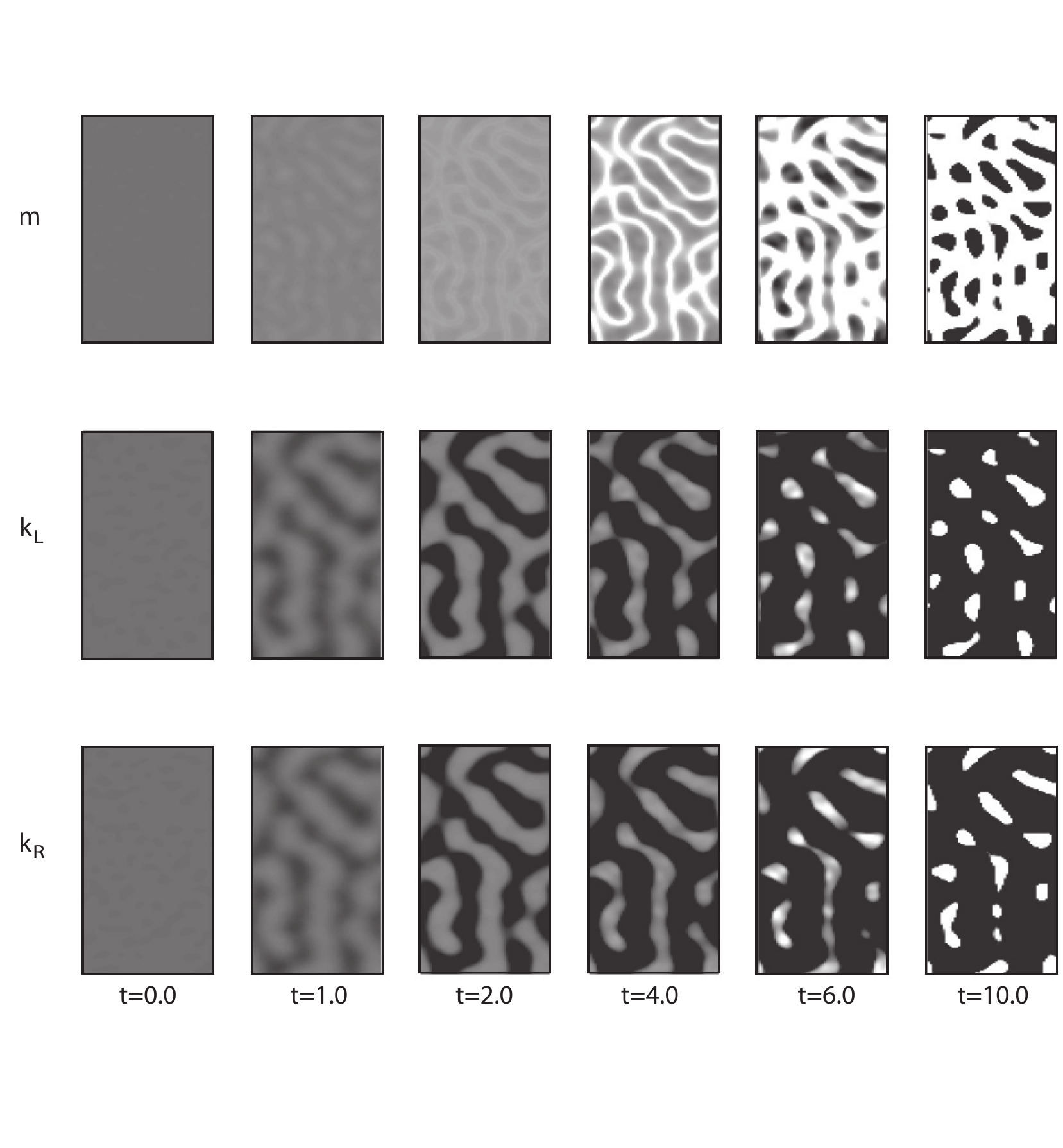}
\end{center}
\vspace{-1.2cm}
\caption{\small Sequence of snapshots showing the progression of $k_L$, $k_R$, and $m$ in layer 2/3 from an initial homogeneous state for the simulation shown in figure \ref{fig:finalPattern}. Note the rapid inheritance of the OD preference of the seed, shown in figure \ref{fig:finalPattern}(a), is evident at $t=2.0$.  Additionally, by $t=6.0$ the koniocellular populations are sequestered away from OD borders that are now purely driven by the M pathway.  The density scale varies from 0 (black) to the maximum local synaptic density  (white). }
\label{fig:timeSeries}
\end{figure}

In this section,  we present numerical results obtained by solving the reduced set of equations (\ref{kL2})--(\ref{m2}). We first generate a pattern of OD columns in layer 4C using the standard correlation--based Hebbian learning rule, equation (\ref{H1}).  An example of such an OD pattern is shown in figure \ref{fig:finalPattern}(a). This then defines the regions $\Sigma_L$ and $\Sigma_R$ or, equivalently, both $w_L$ and $w_R$.  
These, in turn, determine the functions $\chi_{L,R}(\rr)$ in equations (\ref{kL2})--(\ref{m2}). In figure \ref{fig:finalPattern}(b), we plot the OD pattern in layer 2/3 generated from the interlaminar feedfoward projections. Note that the OD properties now vary in a continuous fashion.  Thus, $\Upsilon(\rr) > 1/2$ and $\Upsilon(\rr) <1/2$ if $\rr$ lies above a left-eye column in layer 4C, and vice-versa.  Moreover, $\Upsilon(\rr) \approx 1/2$ near or on an OD border.  The corresponding pattern of $K$ and $M$ afferents in layer 2/3 is shown in figure \ref{fig:finalPattern}(c).  Additionally, we plot the time series evolution of the individual populations for the same simulation in figure \ref{fig:timeSeries}. These numerical simulations confirm the basic predictions of the perturbation analysis presented in section 3. That is, given an initial segregation of left/right M inputs in layer 4C, the vertical inputs from layer 4C to layer 2/3 compete with the koniocellular thalamic inputs $k_L$ and $k_R$.  At each point in layer 2/3, the koniocellular input from the opposite eye to that of the vertical projection from layer 4C rapidly decays to zero, resulting in the projection of layer 4C's OD map onto layer 2/3.  The inclusion of a sampling from layer 4C in the interlaminar projection to layer 2/3 results in the emergence of binocular zones at the OD borders that are fully driven by the magnocellular populations, that is, the koniocellular pathway is restricted to the center of the OD columns.  At the completion of the OD inheritance process, the lateral interactions within layer 2/3 induce a K/M competition that eventually leads to the segregation of the K and M pathway, i.e., the formation of the CO blob lattice.  

We considered a wide array of weight functions and found that an approximate balance of excitation and inhibition performs most robustly.  For example, when excitation strongly dominates the system's sensitivity to homogeneous perturbations is exacerbated, and in the regime where inhibition dominates the growth terms for high wavenumbers do not significantly differ from the growth term corresponding to the critical wavenumber, hence there is a mixing of modes resulting in a highly disorder pattern.  With lateral circuitry that has an approximate balance of excitation and inhibition, our model captures several qualitative aspects of CO blob and OD column formation in superficial layers of primate V1. In particular, it reproduces the experimental observation that the CO blobs are localized around the centers of the columns. In our numerical simulations, we neglected the inhomogeneous terms involving $\psi_L,\psi_R$. This approximation would be valid, for example, under the assumption that the intracortical interactions in the upper laminae are stronger than in the input layer 4C, that is, $|K_1| \ll |K_2|$. It can be checked numerically that the model performs similarly under the inclusion of the $\psi_L,\psi_R$ terms, although it does become more sensitive to the choice of initial conditions.

\begin{figure}[htbp]
\begin{center}
\includegraphics[width=\linewidth]{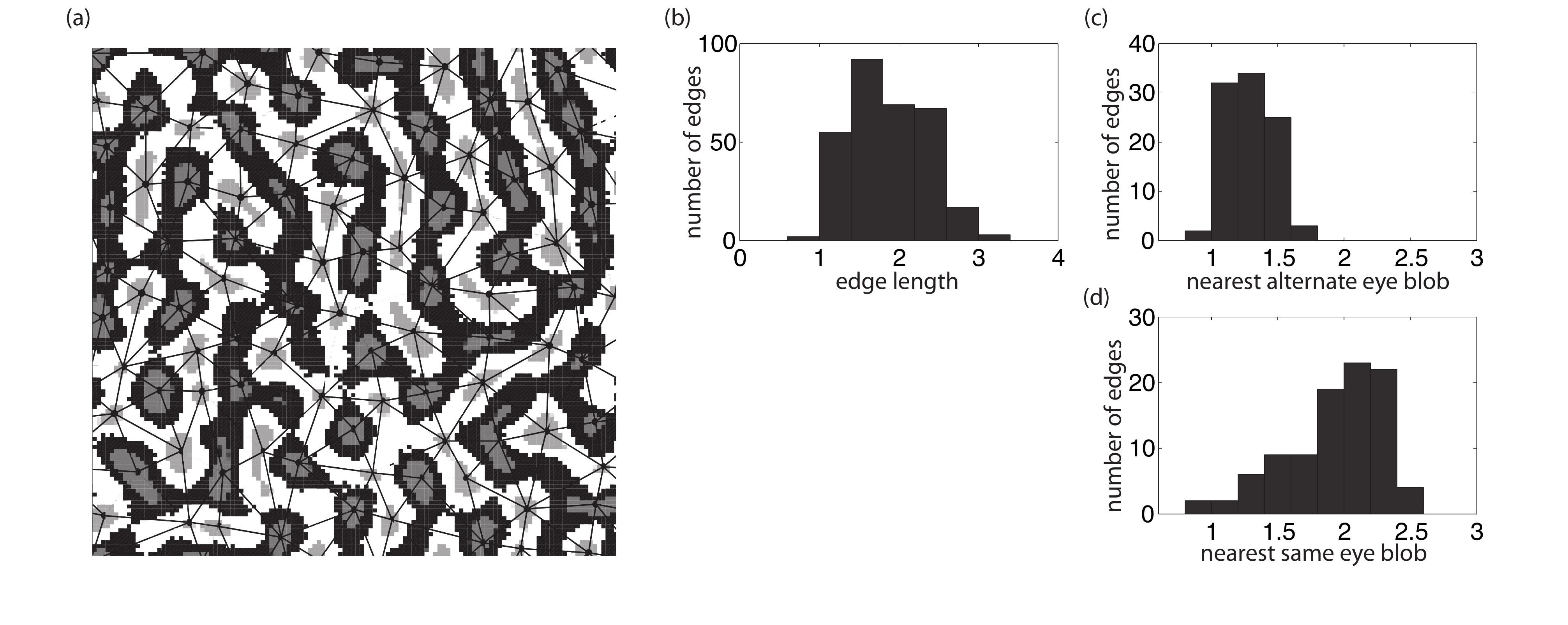}
\end{center}
\vspace{-0.8cm}
\caption{\small (a) Delaunay triangulation of CO blob centers superimposed over an OD/CO blob pattern in layer 2/3 on a 16 x 16 domain. White corresponds to left-eye dominant M input, black corresponds to right-eye dominant M input, and the light and dark grey areas are the left and right eye K inputs, respectively.  (b) Histogram of edge lengths from the triangulation shown in (a).  (c,d) Histograms of the distribution of the nearest neighbor alternate and same eye blobs, respectively. Note the shift in the characteristic distances in these distributions, where the nearest alternate eye blobs are on average more closely packed than the nearest same eye blobs.  The OD pattern matches the layer 4C OD map.  The parameters used are $\widehat{C}_S=C_S=1.0$, $B_S=0.4$, $A_2=8.0$, $\sigma_E=0.42$, $\beta=0.5$, $\sigma_I=0.653$, and $\sigma_f=0.4$.  The initial conditions were $k_0=9.94$ and $m_0=10.0$.}
\label{fig:delaunay}
\end{figure}

As a further comparison with experimental data, we calculate some quantitative aspects of a typical CO blob distribution generated by our model as shown in figure \ref{fig:delaunay}(a).  Following Murphy et al.\   \cite{Murphy:1998:SCO}, we first determine the mean CO blob spacing of the pattern using a Delauney triangulation. We find that the mean edge width of the Delaunay triangulation is approximately twice the mean width of an OD column. Taking the latter to be approximately 385 $\mu m$ \cite{Hubel:1977:FAM}, yields a mean blob spacing of approximately  716 $\mu m$. This is a slight overestimate of the mean value of 590 $\mu m$ found experimentally by Murphy 
{\em et al.\  } \cite{Murphy:1998:SCO}, which could be an artifact of the heightened disorder in the simulated system compared to experimentally found patterns.    However, the Gaussian--like shape of the histogram of edge lengths (figure \ref{fig:delaunay}(b)) is consistent with experimental data.  Another characteristic feature of CO blob distributions in macaque is that the spacing between blobs across OD columns is smaller than between blobs along the same OD column. The former is typically 353 $\mu m$ \cite{Wong-Riley:1984:BCO,Wong-Riley:1994:PVC} whereas the latter is approximately 550 $\mu m$ \cite{Horton:1984:COP}.  An interesting feature of our model is that the profile of the local connectivity in layer 4C determines the periodicity of the OD pattern, whereas the local connectivity in layer 2/3 determines the spacing of the CO blobs within the OD stripes.  Hence, our model can quantitatively account for the asymmetry in CO blob spacing by having different weight functions for the two layers. In figures \ref{fig:delaunay}(c,d), we plot a histogram of the distribution of nearest neighbor alternate and same eye neighbors of the CO blob distribution shown in \ref{fig:delaunay}(b).  It can be seen that the same eye blobs are on average farther away than the alternate eye neighboring blobs, and that both distributions appear roughly as skewed Gaussians, which is consistent with experimental findings.  The ratio of the CO blob spacing along the OD band and across the border is found to be approximately 1.56 in monkeys \cite{Wong-Riley:1984:BCO, Horton:1984:COP}, which is in very good agreement with our simulated distribution that has a ratio of 1.50.  Furthermore, Horton \cite{Horton:1984:COP} noted that there is also an asymmetry in the shape of the blobs themselves.  The blobs are oval shaped of dimension 250 x 150 $\mu m$, which suggests that the ${\mathcal O} (1)$ dynamics determine the length of the minor axis, whereas the spatial interactions within an OD stripe determine the length of the major axis of the CO blob shape.  We do not track statistics for the blob shapes, but note that they have an oval-like profile.  The differences in our model results and experimental observations could be rectified by an alternate choice of the layer 2/3 circuitry and perhaps an alternate initial balance of the K and M pathways.  The experimental data does not suggest a precise choice for these parameters and hence we are only suggesting the underlying structure of the model and representative parameter choices.

\subsection{Monocular Deprivation}

In order to consider the effects of monocular deprivation during development, we relax the assumption that our correlation matrices are symmetric.  In particular, we consider the correlation matrices for the magnocellular(M), koniocellular(K) and correlations between the K and M pathways are respectively given by 
\begin{equation*}
{\mathbf C}=\left ( \begin{array}{cc} C_{LL}  & C_{RL}\\   C_{LR} & C_{RR}  \end{array} \right ),\quad 
\widehat{\mathbf C}=\left (\begin{array}{cc} \widehat{C}_{LL} & \widehat{C}_{RL}\\ \widehat{C}_{LR} & \widehat{C}_{RR}  \end{array} \right
),\quad {\mathbf B}=\left (\begin{array}{cc} B_{LL}  & B_{RL}\\ B_{LR} & B_{RR}  \end{array} \right).
\end{equation*}

The evolution of the synaptic weights is then given by 
\begin{equation*}
\tau_k \frac{dk_L}{dt}= \widehat{C}_{LL} k_L(\rr,t) + \widehat{C}_{RL} k_R(\rr,t) + [ B_{LL} \chi_L(\rr)+B_{RL}\chi_R(\rr)] m(\rr,t) -\eta_0({\mathbf k},m)
\label{kL_MD}
\end{equation*}
\begin{equation*}
\tau_k \frac{dk_R}{dt} = \widehat{C}_{RR} k_R(\rr,t) + \widehat{C}_{LR} k_L(\rr,t) +  [ B_{LR} \chi_L(\rr)+B_{RR}\chi_R(\rr)] m(\rr,t) -\eta_0({\mathbf k},m)
\label{kR_MD}
\end{equation*}
\begin{eqnarray*}
\tau_m \frac{dm}{dt} &=&  \biggr ( {C}_{LL}\chi_L^2(\rr) +{C}_{RR}\chi_R^2(\rr) + [C_{LR}+C_{RL}] \chi_L(\rr) \chi_R(\rr) \biggr )   m(\rr,t) \nonumber \\ 
    & &  + \left (B_{LL}\chi_L(\rr)+B_{RL}\chi_R(\rr) \right ) k_L(\rr,t)  \nonumber \\ 
    & & + \left (B_{LR}\chi_L(\rr)+B_{RR} \chi_R(\rr)\right ) k_R(\rr,t) -\eta_0({\mathbf k},m),
\label{m_MD}
\end{eqnarray*}
where $\eta_0$ is the corresponding subtractive normalization term and $\chi_{L, R}$ are defined in equation (\ref{chiJ}). 

We consider normal development until MD onset, then reduce the correlation coefficients associated with the deprived eye, similar to the approach in \cite{Miller:1989:ODC} to simulate MD.  At the initiation of MD, there is an approximate balance of left/right eye dominated regions with CO blobs centered within the OD stripes.  Below in figure \ref{fig:MD}, we display the simulated effects of MD on the OD pattern and distribution of the CO blobs. Without any loss of generality, we take the right eye to be deprived (black regions) while the left eye corresponds to the non-deprived eye (white regions), and CO blobs are colored in gray.  At the onset of MD, the correlation coefficients corresponding the right eye activity driving cortical activity are reduced (as in \cite{Miller:1989:ODC}.   MD results in the broadening of the OD stripes associated with the non-deprived eye broaden and a reduction of the OD stripes associated with the deprived eye, while there are minimal alterations of the CO blob positions.  This numerical study is inline with the experimental finding of Murphy {\em et al.\  }  \cite{Murphy:2001:COB}.  

\begin{figure}[htbp]
\begin{center}
\includegraphics[width=0.7\linewidth]{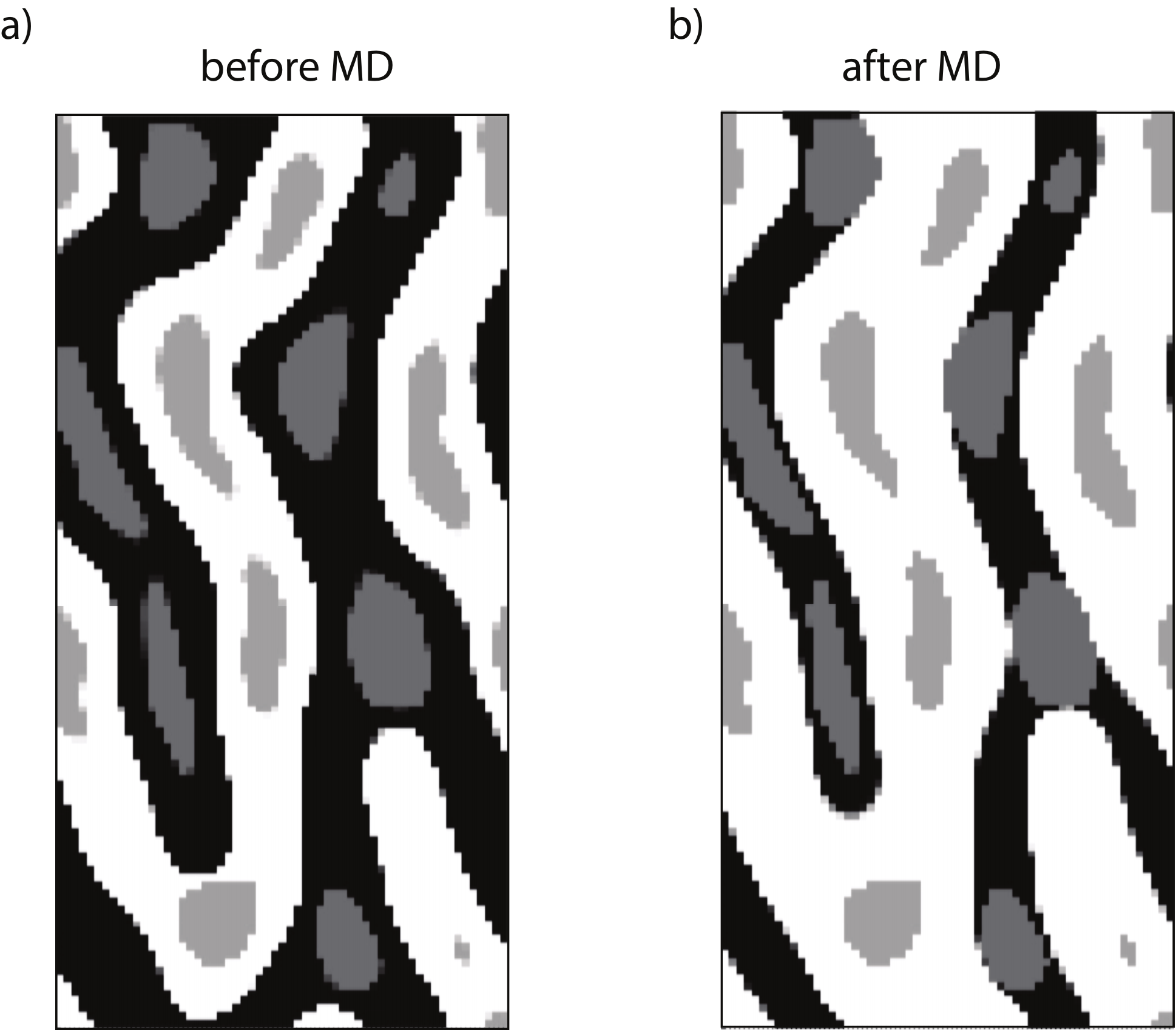}
\end{center}
\caption{\small Effects of monocular deprivation on the ocular dominance pattern and CO blob distribution.  Here black corresponds to the deprived eye, whereas white indicates the remaining open eye, while CO blob regions are colored in gray.  In (a), the OD pattern and CO blob distribution immediately before the onset of MD is shown, whereas in (b) the aftereffects of MD is pictured. Before deprivation the non-deprived eye area constituted 48.2\% of region, while post deprivation the  non-deprived controlled 69.3\% of the cortical region.  Note that while the OD pattern is significantly modified, the CO blob distribution remains preserved.  The domain size is 6 $\times$ 12 and is meant to simulate a 2.3 $mm$ by 4.6 $mm$ region of cortex. The parameters used in the simulation prior to MD are identical as those in figure 7 with the exception that $\epsilon = 0.06$ and $C_d = 0.2$, while during MD we take  $C_{LL} = \widehat{C}_{LL} =  1$, $B_{LL} =  0.4$, $\widehat{C}_{RR} =  0.2$, $C_{RR} =  0.3,$ $ B_{RR} =  0.08$, $\widehat{C}_{LR} = \widehat{C}_{RL} =0$, $C_{LR} =  0.2$, $C_{RL} =  0.067$, and $B_{LR} = B_{RL} = 0 $.      }
\label{fig:MD}
\end{figure}


\section{Discussion}

In this paper, we constructed a multi-layer, activity-dependent model for the joint development of ocular dominance columns and cytochrome oxidase blobs in primary visual cortex of primates. We assumed that the OD map first forms in layer 4C and is then inherited by the supragranular layers via feedforward vertical projections from layer 4C.  Competition between these feedforward projections and the direct thalamic input to layer 2/3 then resulted in the formation of CO blobs superimposed upon the OD map. The CO blob distribution obtained from numerical simulations was consistent with experimental data. First, the CO blobs were located in the centers of OD columns. In our model this was related to the fact that the feedforward projections sampled over a local region of layer 4C resulting in binocular regions along the OD borders. Second the CO blob lattice exhibited a characteristic asymmetry, in which the spacing of a CO blob from the nearest neighbor blob driven by the same-eye was on average greater than the spacing from the nearest neighbor blob driven by the alternate-eye. In our model this was a consequence of the fact that the periodicity of the OD map that develops in layer 4C was determined via the lateral circuitry in layer 4C, whereas the periodicity of the CO blob lattice within an OD stripe was determined via the circuitry in layer 2/3. For simplicity, we used a linear correlation-based Hebbian rule with subtractive normalization. The advantage of a linear model is that one can explicitly determine the propagation of weight modifications through the layers. However, one of the possible limitations of such a model is the sensitivity to initial conditions. It might be possible to construct a more robust model by considering a nonlinear competitive Hebbian learning rule \cite{Piepenbrock:2000:EIC}, although the analysis would be considerably more involved. 

As a first approximation we considered a purely feedforward model in this paper. This was partly motivated by the observation that there are no direct feedback connections from layer 2/3 to layer 4C. However, connections between layer 5 and layer 6 could provide a more indirect feedback pathway for activity in layer 2/3 to modulate the activity in layer 4C.  Furthermore, it has been demonstrated in cat that at the onset of monocular deprivation (MD), plasticity first occurs at the OD borders in layer 2/3 and only after a delay of several days does any modification in the direct thalamic input to layer 4 occur, suggesting that layer 2/3 influences the development of layer 4 during the critical period \cite{Trachtenberg:2001:RAP}. Studies in mice indicate that layer 2/3 plasticity induced by monocular deprivation involves a rapid restructuring of dendritic spines that is regulated by GABA inhibition (reviewed in \cite{Hensch:2005:CPP}).  Hence, just as in early development when OD columns and CO blobs first form, it is likely that the laminar structure of cortex plays an important role during the critical period. In particular, one cannot model monocular deprivation simply as a change in input correlations to a single layer network. We hope in future work to extend our laminar network model to analyze development during the critical period, and to understand the delayed occurrence of plasticity in layer 4 during monocular deprivation by considering the role of feedback from layer 2/3 to layer 4. Finally, note that the principles outlined in this paper could be extended to include orientation selectivity and spatial frequency in the visual system, and could also be employed to study the development of laminar structures in other cortical areas.  


\vspace{-3ex}
\begin{backmatter}

\section*{Competing interests}
  The author declares no competing interests.

\section*{Acknowledgements}
This work was commenced during my graduate study mentored by Paul C.\ Bressloff (PCB).  Discussions with PCB and insights from PCB have been invaluable.  This work was partially supported by RTG 0354259. 
 

\bibliographystyle{bmc-mathphys} 
\bibliography{bibLaminar}


\end{backmatter}

\end{document}